\documentclass[aps,prl,reprint,showpacs,superscriptaddress]{revtex4-1}
\UseRawInputEncoding
\usepackage{graphicx}
\usepackage{color}
\usepackage[colorlinks, linkcolor=blue, anchorcolor=blue, citecolor=blue, urlcolor=blue]{hyperref}
\usepackage{multirow}
\usepackage{amsmath}

\begin{document}


\title{Gapless triangular-lattice spin-liquid candidate in PrZnAl$_{11}$O$_{19}$}



\author{Huanpeng Bu}
\affiliation{Neutron Science Platform, Songshan Lake Materials Laboratory, Dongguan, Guangdong 523808, China}

\author{Malik Ashtar}
\affiliation{School of Physics and Wuhan National High Magnetic Field Center,
Huazhong University of Science and Technology, Wuhan 430074, P. R. China.}

\author{Toni Shiroka}
\affiliation{Laboratory for Muon Spin Spectroscopy, Paul Scherrer Institute, CH-5232 Villigen PSI,
Switzerland}
\affiliation{Laboratorium f\"{u}r Festk\"{o}rperphysik, ETH Z\"{u}rich, CH-8093 Zurich, Switzerland}

\author{Helen C. Walker}
\affiliation{ISIS Neutron and Muon Source, Rutherford Appleton Laboratory, Chilton, Didcot OX11 0QX, United Kingdom}

\author{Zhendong Fu}
\affiliation{Neutron Science Platform, Songshan Lake Materials Laboratory, Dongguan, Guangdong 523808, China}

\author{Jinkui Zhao}
\affiliation{Neutron Science Platform, Songshan Lake Materials Laboratory, Dongguan, Guangdong 523808, China}
\affiliation{Institute of Physics, Chinese Academy of Sciences, Beijing 100190, China}

\author{Jason S. Gardner}
\affiliation{Material Science \& Technology Division, Oak Ridge National Laboratory, Oak Ridge, Tennessee 37831, USA}

\author{Gang Chen}
\affiliation{Department of Physics and HKU-UCAS Joint Institute for Theoretical
and Computational Physics at Hong Kong, The University of Hong Kong, Hong Kong, China}

\author{Zhaoming Tian}
\email{tianzhaoming@hust.edu.cn}
\affiliation{School of Physics and Wuhan National High Magnetic Field Center,
Huazhong University of Science and Technology, Wuhan 430074, P. R. China.}

\author{Hanjie Guo}
\email{hjguo@sslab.org.cn}
\affiliation{Neutron Science Platform, Songshan Lake Materials Laboratory, Dongguan, Guangdong 523808, China}



\date{\today}

\begin{abstract}
  A quantum spin liquid (QSL) is an exotic state in which electron spins are highly entangled, yet keep fluctuating even at zero temperature. Experimental realization of model QSLs has been challenging due to imperfections, such as antisite disorder, strain, and extra or a lack of interactions in real materials compared to the model Hamiltonian. Here we report the magnetic susceptibility, thermodynamic, inelastic neutron scattering (INS), and muon-spin relaxation studies on a polycrystalline sample of PrZnAl$_{11}$O$_{19}$, where the Pr$^{3+}$ ions form an ideal two-dimensional triangular lattice. Our results demonstrate that this system does not order nor freeze, but keep fluctuating down to 50 mK despite large antiferromagnetic couplings ($\sim$ -10 K). Furthermore, the INS and specific-heat data suggest that PrZnAl$_{11}$O$_{19}$ is best described as a gapless QSL.
\end{abstract}


\maketitle


\section{Introduction}

Frustration, which arises when different interactions cannot be minimized simultaneously, is ubiquitous in condensed matter physics. In magnetic materials, frustration suppresses the formation of a long-range magnetically ordered state. In some cases, the ordering is suppressed even down to zero Kelvin, but the spins remain highly entangled over long distances. Such a ground state, known as a quantum spin liquid (QSL) state, is highly degenerate and sensitive to perturbations. A QSL can host exotic properties such as fractional excitations, which have the potential for application in quantum computations if braided properly \cite{Nayak2008}. It is also intimately connected to high-temperature superconductors as pointed out by Anderson \cite{Anderson1987}, who first proposed this intriguing state based on an \textit{S} = 1/2 triangular lattice with nearest-neighbor Heisenberg antiferromagnetic interactions \cite{Anderson1973}. QSLs have been intensively investigated, both theoretically and experimentally \cite{Balents2010,Savary2016,Zhou2017,Broholm2020}, and different kinds of QSLs have been proposed and classified according to their symmetries \cite{Wen2002}. One important feature that distinguishes the different classes of QSL is whether the excitation is gapped or gapless with power law spin-spin correlations \cite{Broholm2020}.

Experimentalists are endeavouring to realize this intriguing ground state based on geometrically frustrated lattices such as the two-dimensional (2D) triangular lattice \cite{Shimizu2003,Yamashita2008}, kagome lattice \cite{Helton2007}, three-dimensional pyrochlore lattice \cite{Gardner1999,Ross2011,Kimura2013}, and more recently the honeycomb lattice \cite{Jackeli2009,Plumb2014,Banerjee2016}. Most of these studies have focused on magnetic ions with a small quantum number, such as Cu$^{2+}$ with \textit{S} = 1/2, to enhance the quantum fluctuations. On the other hand, ions with a large spin-orbit coupling, combined with crystal-electric-field (CEF) effects, may also result in marked quantum effects due to the formation of an effective spin-1/2 state alongside the anisotropic magnetic interactions, such as in 4\textit{d} or 5\textit{d} systems \cite{Plumb2014,Banerjee2016}, or in 4\textit{f} rare earth systems \cite{Li2015,Li2016,Shen2016,Shen2018,Ding2019,Dai2021}. The insulating 4\textit{f} electron systems are of particular interest since the electrons are more localized, and the exchange interactions are more short-ranged compared to that of the \textit{d} electrons, thus simplifying the model Hamiltonian.

Real materials always suffer from impurities and/or disorder, which can have a profound impact on the properties of the QSL. For example, antisite disorder between Zn and Cu is expected in the kagome herbertsmithite ZnCu$_3$(OH)$_6$Cl$_2$ \cite{Freedman2010}, where the inter-layer Cu- and/or the Zn ions within the kagome lattice may introduce spin-exchange randomness and influence the low-energy excitations significantly \cite{Singh2010}. Even when the disorder is outside the magnetic layers, such as in triangular YbMgGaO$_4$ where Mg and Ga ions exchange sites, it can lead to a spin-glass state, and even be responsible for spinon-like excitations \cite{Ma2018}. However, disorder is not always harmful to the QSL, since, under certain circumstances, it can facilitate quantum fluctuations \cite{Ross2009,Chang2012,Furukawa2015}.

Another class of 2D triangular frustrated magnet based on rare-earth ions, \textit{R}ZnAl$_{11}$O$_{19}$ (\textit{R} = rare earth), was reported recently \cite{Ashtar2019}. One advantage of this series of compounds is that the ionic radii of the magnetic and nonmagnetic ions differ significantly, e.g., 1.126 \AA~for Pr$^{3+}$, 0.6 \AA~for Zn$^{2+}$, and 0.535 \AA~for Al$^{3+}$. Thus, the site mixing between magnetic and nonmagnetic ions is not possible. Moreover, the Pr triangular layers are separated by \textit{c}/2 $\sim$ 11.0 \AA, making it close to an ideal 2D structure, and again minimizing disorder effects (if any) outside the magnetic layers.  As a comparison, the interlayer distance is about 8.4 \AA~for YbMgGaO$_4$ \cite{Li2019}. Therefore, this series of compounds seem to show potential for
hosting exotic ground states considering the high Curie-Weiss temperature and the lack of magnetic ordering down to 0.43 K \cite{Ashtar2019}.

In this paper, we deepen our understanding of the spin dynamics and low-energy excitations of PrZnAl$_{11}$O$_{19}$ by utilizing ac susceptibility, specific-heat, inelastic neutron scattering (INS), and muon-spin relaxation measurements on polycrystalline samples. No magnetic ordering or spin freezing was detected down to 50 mK. Instead, substantial gapless low-energy magnetic excitations were revealed by specific-heat and INS measurements. The low-energy diffusive excitations, together with a $T^2$ behavior of the specific heat at low temperatures point to the emergence of a gapless QSL state. Our data also reveal a peculiar temperature dependence of the specific heat in a magnetic field, which deviates from the $T^2$ behavior at modest fields, and recovers again above 9 T.

\section{Experiment}

Polycrystalline samples of PrZnAl$_{11}$O$_{19}$ were prepared using a standard solid-state reaction technique. Raw materials of Pr$_6$O$_{11}$ (99.99\%), ZnO(99.99\%), and Al$_2$O$_3$ (99.99\%) were dried at 900$^\circ$C over night prior to reaction to avoid moisture contamination. Then, the starting materials were mixed in the stoichiometric ratio and ground thoroughly using an agate mortar, pressed into pellets, and calcined at 1550$^\circ$C for 5 days with several intermediate grindings.
The phase purity of the sample was confirmed by X-ray powder diffraction (XRD) measurement with Cu $K_\alpha$ radiation.

The dc and ac magnetic susceptibility between 2 and 350 K were measured using the vibrating sample magnetometer (VSM) and ACMS-II options, respectively, of the physical property measurement system (PPMS DynaCool, Quantum Design). Sub-Kelvin ac susceptibility and heat capacity measurements were carried out with a dilution insert of the PPMS. For the ac susceptibility measurement, a driven field of 1-3 Oe in amplitude was used.

Inelastic neutron scattering (INS) measurements were performed on the MERLIN spectrometer at ISIS, UK. The samples were loaded into aluminium foil sachets, which were wrapped around the inside of a cylindrical aluminium can and cooled down to 7 K by a close-cycled refrigerator. MERLIN was operated in multi-rep mode scattering neutrons with incident energies of 23.0, 36.5 and 67.1 meV. Data \cite{data1,data2} were processed using Mantid, and the phonon signal were removed from the Pr sample data using the isostructural non-magnetic La sample data appropriately scaled for relative sample masses.

Zero-field (ZF) and longitudinal-field (LF) muon-spin relaxation ($\mu$SR) measurements were performed on the Dolly spectrometer at the Paul Scherrer
Institute (PSI), Villigen, Switzerland. Nearly 100\% polarized muons were injected into the sample
and the decay positrons, which are emitted preferentially along the muon spin direction, were
detected. The asymmetry is defined as $A(t) = [N(t) - \alpha B(t)]/[N(t) + \alpha B(t)]$, where $N(t)$
and $B(t)$ are the number of positrons hitting the forward and backward detectors at time \textit{t}, while the parameter $\alpha$ reflects the relative counting efficiency of the two detectors.

\section{Results and discussions}

\begin{figure}
  \centering
  \includegraphics[width=0.9\columnwidth]{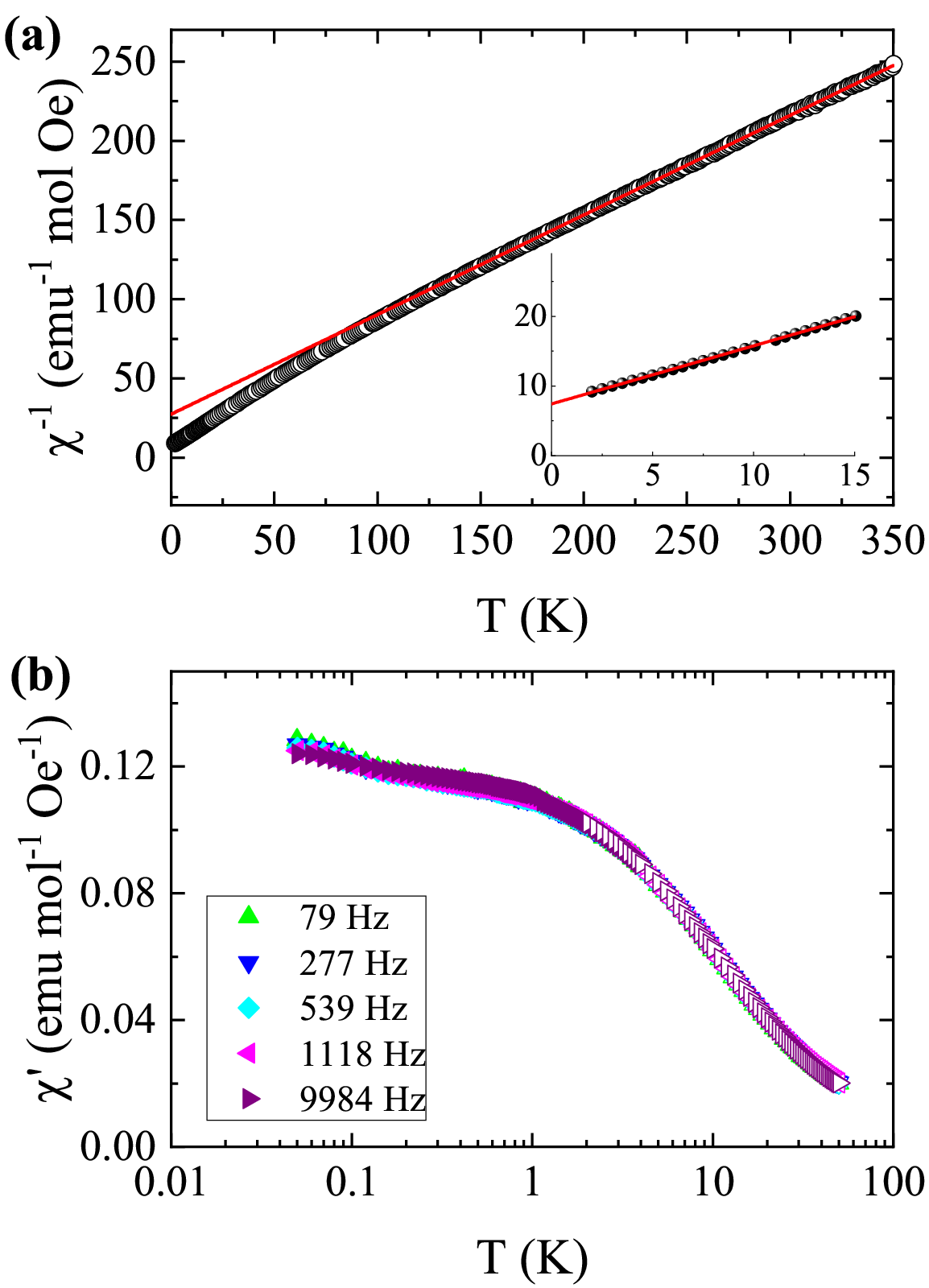}\\
  \caption{(a) Temperature dependence of the inverse dc magnetic susceptibility, $\chi^{-1}$ measured in a small magnetic field of 100 mT. The red line is a fit to the Curie-Weiss law. The inset shows the low-temperature region, where a CW-like fitting is performed. (b) Temperature dependence of the real component of the ac susceptibility, $\chi'$, measured at various frequencies. The open- and closed symbols represent data obtained using the ACMS-II and ACDR options, respectively. No frequency dependent behavior can be observed in the whole temperature range.}\label{sus}
\end{figure}

Figure \ref{sus}(a) shows the temperature dependence of the inverse magnetic susceptibility $\chi^{-1}$. No sign of magnetic ordering is observed down to 2 K. The data above 200 K can be well fitted to the Curie-Weiss (CW) law $\chi$ = C/(\textit{T}-$\theta_{CW}$), which yields an effective moment $\mu_{eff}$ of 3.57 $\mu_B$/Pr and a Curie-Weiss temperature $\theta_{CW}$ of -44 K. Below $\sim$ 100 K, the susceptibility deviates from CW behavior, most likely due to the crystal-electric-field (CEF) effect. Therefore, a CW-like fit to the linear region below 15 K, which results in a negative $\theta_{CW}$ of -8.9 K, provides another measure of the interaction strength and agrees well with previous studies \cite{Ashtar2019}. In addition, we probe the spin dynamics down to 50 mK using ac susceptibility measurements.
As shown in Fig. \ref{sus}(b), the susceptibility $\chi'$ increases monotonically with decreasing temperature, and tends to level off below $\sim$1 K with a large value, indicating substantial low-energy excitations. Moreover, it shows a frequency-independent behavior, ruling out any spin freezing or spin-glass transition down to 50 mK. This clearly demonstrates that the spins keep fluctuating down to 50 mK, despite a large negative Curie-Weiss temperature of -9 K, which results in a large frustration index (\textit{f} $>$ 9/0.05 = 180).

\begin{figure}
  \centering
  \includegraphics[width=0.9\columnwidth]{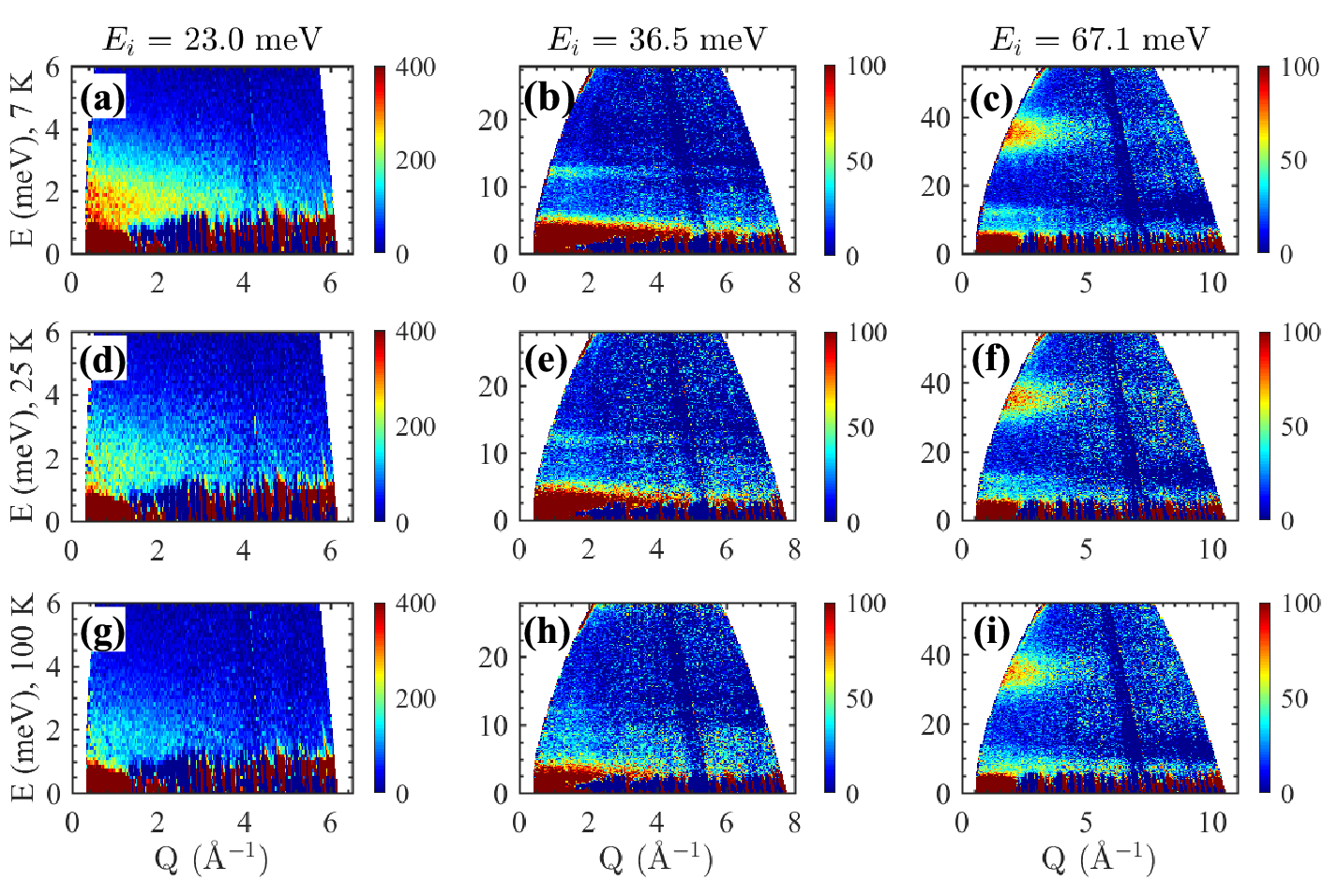}\\
  \caption{INS intensity maps with different $E_i$s from the PrZnAl$_{11}$O$_{19}$ sample. The phonon contributions have been subtracted using a LaZnAl$_{11}$O$_{19}$ reference sample. The maps were obtained at 7 K (a-c), 25 K (d-f), and 100 K (g-i).}\label{INS}
\end{figure}

\begin{figure}
  \centering
  \includegraphics[width=0.9\columnwidth]{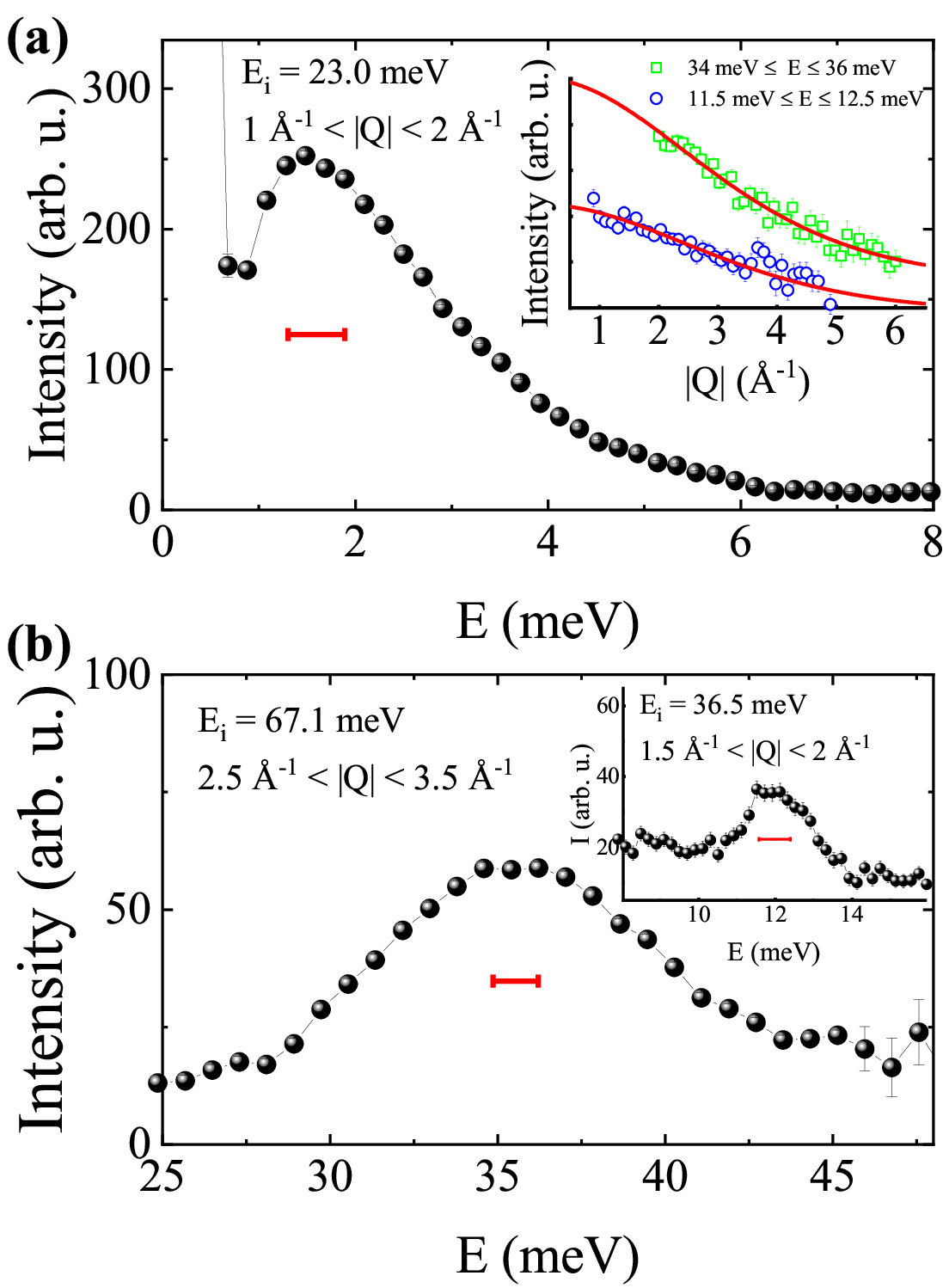}\\
  \caption{Energy cuts at different $|\mathrm{Q}|$ positions. The red bars indicate the instrument energy resolution at the specific E position. The inset of (a) shows the $|\mathrm{Q}|$ dependence of the intensity for the 12- and 36-meV excitations, which follow the magnetic form factor of Pr$^{3+}$, $|f(Q)|^2$, plus a small constant background.}\label{cut}
\end{figure}

Pr$^{3+}$ (4$f^2$, J = 4) is a non-Kramers ion with an even number of electrons per site. Under the $D_{3h}$ symmetry, the degenerate nine-fold multiplet is split into three singlets and three doublets. In order to determine the CEF scheme and identify any low-energy excitations, we performed INS measurements at MERLIN \cite{MERLIN}, ISIS. As shown in Fig. \ref{INS}(a-c), two dispersionless excitations can be observed at \textit{E} $\sim$ 12 and 36 meV. The Q dependence of the integrated intensities follows the magnetic form factor of Pr$^{3+}$, as shown in the inset of Fig. \ref{cut}(a), confirming their magnetic origin. Upon increasing the temperature, the intensities of these two excitations decrease, and almost disappear at 100 K for the 12-meV excitation. These observations suggest that these are the CEF excitations.
At 100 K, the 12-meV crystal-field level has been thermally populated at the expense of the ground state, resulting in the weak intensity of the 12-meV excitation at this temperature.
It is obvious that the excitation at 36 meV is much broader in energy compared to that at 12 meV. This can be seen more easily from the constant Q cuts as shown in Fig. \ref{cut}(b) and the inset. While the peak width of the 12-meV excitation is comparable to the instrument resolution ($E_i$ = 36.5~meV, $E$ = 12.0~meV, $\Delta E_\mathrm{inst}$ = 0.8~meV), it is much broader for the 36-meV excitation ($E_i$ = 67.1~meV, $E$ = 36.0~meV, $\Delta E_\mathrm{inst}$ = 1.4~meV). Thus, two or more near degenerate excitations around 36 meV can be expected.

The most prominent feature in Fig. \ref{INS} is the diffusive low-energy excitations (at $\sim$1.5 meV) with a substantial spectral weight at low Q
at low temperatures (7 K).
The constant Q cut, as shown in Fig. \ref{cut}(a), shows a distinct peak profile compared to that of the other two excitations. As can be seen, the peak is asymmetric, with a long tail at the high-energy side, which is reminiscent of the excitation continuum due to spinons observed in some QSL candidates \cite{Shen2018,Plumb2019}. On the contrary, the peaks at 12 and 36 meV have a more symmetric profile. Also, the peak width is much larger than the instrument resolution ($E_i$ = 23.0~meV, $E$ = 1.5~meV, $\Delta E_\mathrm{inst}$ = 0.6~meV). We exclude the CEF origin for these excitations, as will be discussed later for the magnetic entropy.

In summarizing the INS data, we observe two excitations at $\sim$12 and 36 meV which behaves like CEF excitations, with possible overlapping of multiple CEF levels at $\sim$36 meV.
The 12-meV CEF level is consistent with the crossover temperature around 100 K in the magnetic susceptibility.
According to the point symmetry, there are six CEF levels, and thus
one could expect to observe up to five excitations from the ground state at base temperature.
Here we measured the excitations up to an energy transfer of 135 meV, but no further excitations could be identified above 36 meV. The current data set is insufficient for us to rule out possible CEF transitions at higher energies, and we are unable to determine the full CEF scheme and the corresponding CEF wave functions at the moment. However, this does not influence our main conclusion on the dynamic nature of the material, since the electrons prefer to occupy the low-energy CEF levels at low temperatures.

\begin{figure*}
  \centering
  \includegraphics[width=2\columnwidth]{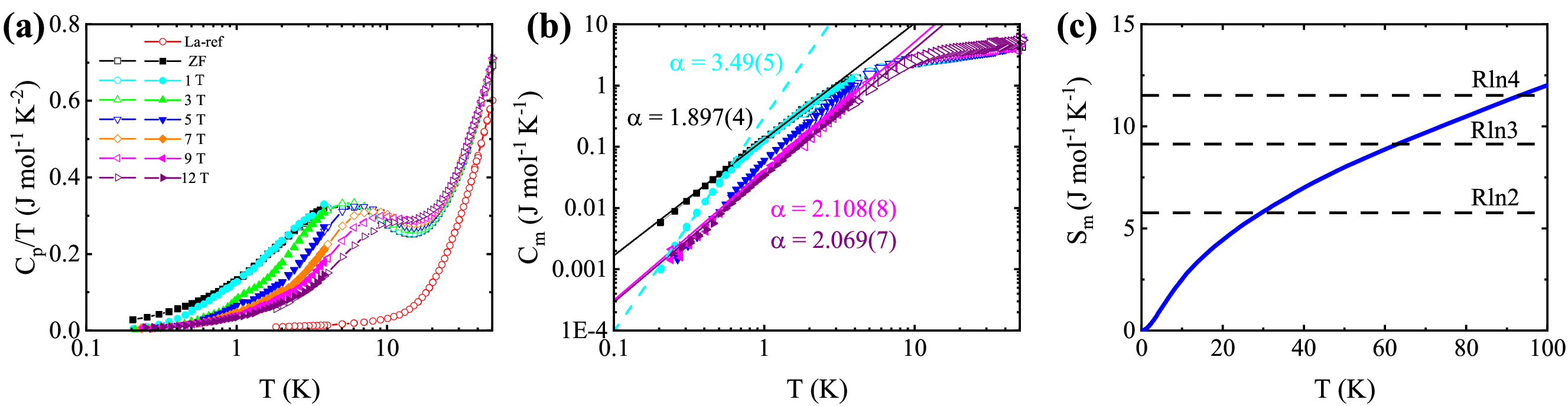}\\
  \caption{(a) Specific heat of PrZnAl$_{11}$O$_{19}$ measured at various magnetic fields. The open and closed symbols represent the data sets obtained in the He4 and dilution-refrigerator temperature regimes. The phonon contribution obtained from the renormalized LaZnAl$_{11}$O$_{19}$ data is also shown. (b) Temperature dependence of the magnetic specific heat in a log-log scale. The lines are fits according to a power law, whose exponents are also shown. (c) The ZF magnetic entropy is obtained by integrating the magnetic specific heat $C_m/T$.}\label{HC}
\end{figure*}

The low-energy excitations were further probed by specific-heat measurements.
As shown in Fig. \ref{HC}(a), only a broad peak at $\sim$ 5 K could be observed in ZF, indicating no long-range magnetic ordering.
The peak is suppressed by a magnetic field, and shifted to higher temperatures with increasing fields. In order to obtain the magnetic contributions, we measured an isostructural reference LaZnAl$_{11}$O$_{19}$, whose signal was renormalized taking into account the atomic mass difference \cite{Bouvier1991}, and then subtracted from the total specific heat. The obtained magnetic specific heat, $C_m$ is shown in Fig. \ref{HC}(b).
We note that $C_m$ cannot be described by a multi-level Schottky anomaly, as usually observed for rare-earth ions with CEF splitting. Instead, it shows a clear power law behavior as $C_m$ = A$T^{\alpha}$, indicating a gapless excitation. This corroborates our conclusion from the INS results that the excitation around 1.5 meV is not a CEF excitation. The fit to the ZF data below 2 K yields $\alpha$ = 1.897(4). Such a quasi-quadratic behavior would be consistent with a Dirac QSL state in which a $C_m \propto$ $T^2$ behavior due to the Dirac nodes is expected.
A $T^2$ specific heat in two dimensions has also been obtained in other frustrated magnets such as
the spin-1 triangular lattice antiferromagnet NiGa$_2$S$_4$~\cite{Nakatsuji2005,Podolsky2009,Stoudenmire2009}
and the spin-2 triangular lattice antiferromagnet FeAl$_2$Se$_4$~\cite{Li2019prb}.
The quasi-quadratic specific heat in PrZnAl$_{11}$O$_{19}$ here should be fundamentally different from the ones in NiGa$_2$S$_4$ and FeAl$_2$S$_4$.
In NiGa$_2$S$_4$ and FeAl$_2$S$_4$, it was attributed to the emergent gapless Halperin-Saslow mode and glassy-like freezing
that result from the non-magnetic disorder and the continuous spin-rotational symmetry~\cite{Podolsky2009,Stoudenmire2009}.
For the Pr triangular lattice in PrZnAl$_{11}$O$_{19}$, due to the spin-orbit coupling, the
effective model between the non-Kramers doublets is highly anisotropic~\cite{Liu2018},
and there is no such continuous symmetry breaking.
Taking into account the above argument and the absence of spin freezing, we thus
propose that the disordered state in PrZnAl$_{11}$O$_{19}$ is more likely to be a Dirac QSL.

Interestingly, the specific heat shows an unusual magnetic-field dependence. When a field of 1 T is applied, $C_m$ shows a crossover behavior between different temperature regimes. Between 0.5 and 2 K, the quasi-quadratic behavior remains. Below 0.5 K, however, it still follows the power law, but with a power of 3.49(5), as shown in Fig. \ref{HC}(b). Such a separation is well defined up to 3 T. Above 9 T, the quadratic behavior is recovered, with $\alpha$ = 2.108(8) and 2.069(7) for the 9 and 12 T data sets, respectively. This field dependent behavior is in contrast with the one predicted for the Dirac QSL, for which a linear \textit{T} dependence is often expected \cite{Ran2007,Zeng2022}. On the other hand, a spinon Fermi surface U(1) QSL is predicted to exhibit a $T^{2/3}$ behavior in zero field \cite{Motrunich2005}, although a linear $T$ behavior is usually observed experimentally \cite{Dai2021}. It is, however, important to notice that we are measuring a polycrystalline sample.
Due to this fact, the effective magnetic field experienced by the dipole component of the local Pr non-Kramers moments in each grain
depends on the orientation of the grain crystallographic axes. Thus, the actual magnetic field is not uniform throughout the sample, and
we are faced with the possibility of random fields. A combination of random fields with a precise microscopic spin model is needed to further
analyze and understand the unusual magnetic-field dependence of the specific heat.

The ZF magnetic entropy is obtained by integrating $C_m$/T from the base temperature and shown in Fig. \ref{HC}(c). The entropy increases smoothly with increasing temperature, showing no noticeable plateau, or significant release of entropy, indicative of a phase transition. At 30 K, the released entropy is almost equal to \textit{R}ln2, where \textit{R} = 8.314 J~mol$^{-1}$~K$^{-1}$ is the ideal-gas constant. The observation of magnetic responses in the ac susceptibility down to 50 mK indicates that the ground state is a non-Kramers doublet. Even if the ground state is not a non-Kramers doublet, and the magnetism originates from the Van Vleck paramagnetism due to low-lying singlets, the gap between the singlets should be so small that they can be considered as a quasi-doublet at 50 mK. An earlier ESR study reveals that the ground state doublet is anisotropic as characterized by two distinct Land\'{e} g factors \cite{Ashtar2019}. The diffusive excitations around 1.5 meV could also be overlaps of some CEF levels, as observed in low symmetric Pr$^{3+}$ compounds such as PrNiSn \cite{McEwen2006}. However, accepting this doublet ground state, and consider the released entropy at 30 K, it is difficult to model a complex scheme of CEF levels around 1.5 meV ($\sim$17 K), but rather it is more appropriate to ascribe it to a gapless continuum due to spinons. At higher temperatures, the phonon subtraction using a non-magnetic reference sample could be subject to some uncertainties. Therefore, we calculate the entropy up to 100 K. The obtained entropy at 100 K is roughly equal to \textit{R}ln4, so that the CEF excitation observed by INS at 12 meV is likely a doublet too, or an overlap of two singlets. Note, that the ground-state doublet is protected by the crystal symmetry, rather than the time-reversal symmetry, so that it could be lifted due to potential Jahn-Teller distortions. However, powder XRD measurements down to 12 K (data not shown) do not indicate any lowering of the crystal symmetry with respect to high temperature.

\begin{figure}
  \centering
  \includegraphics[width=0.9\columnwidth]{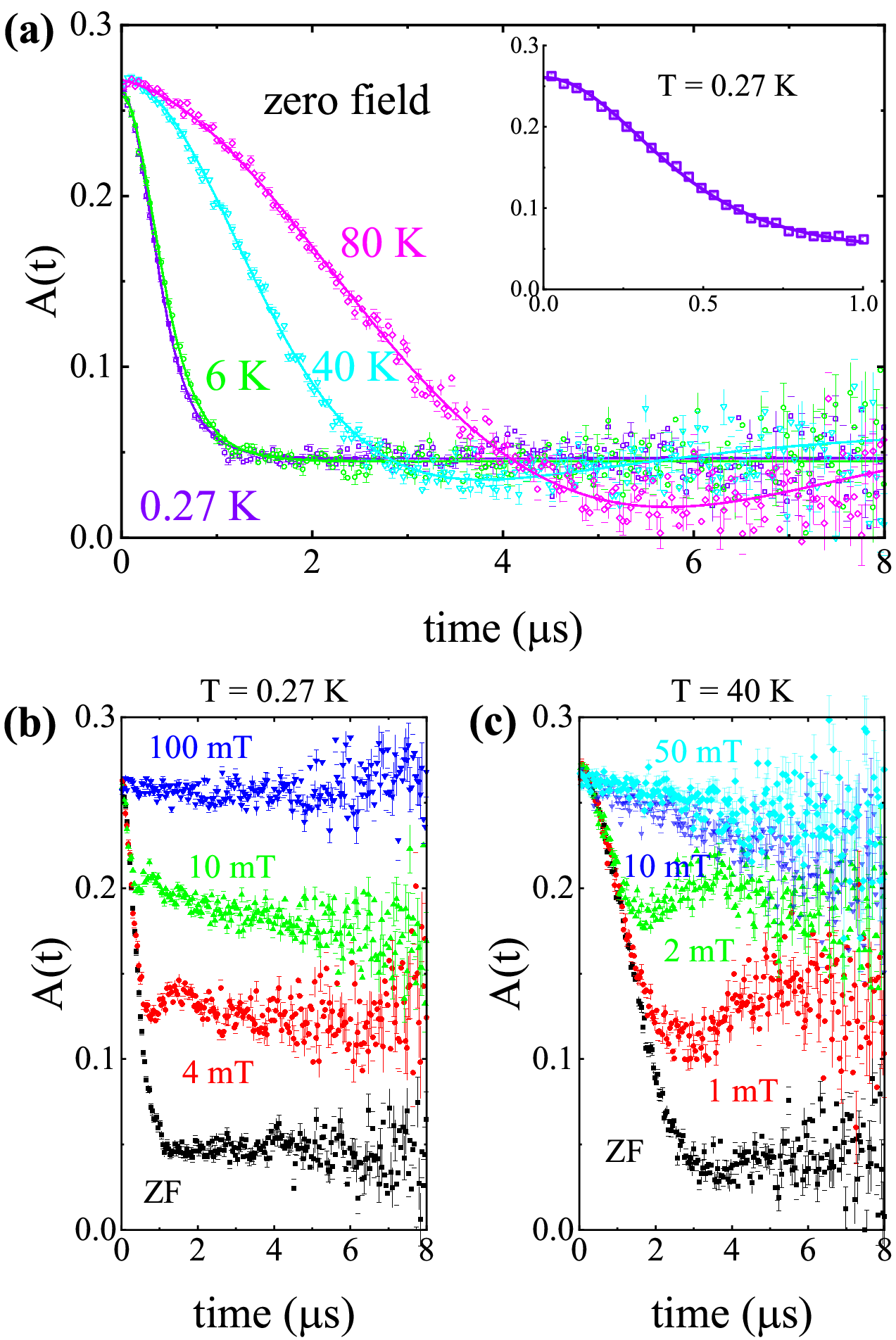}\\
  \caption{(a) ZF-$\mu$SR spectra measured at various temperatures. The solid curves represent the fit as described in the text. (b) and (c) LF-$\mu$SR spectra measured at 0.27 and 40 K, respectively.}\label{musr}
\end{figure}

\begin{figure}
  \centering
  \includegraphics[width=0.9\columnwidth]{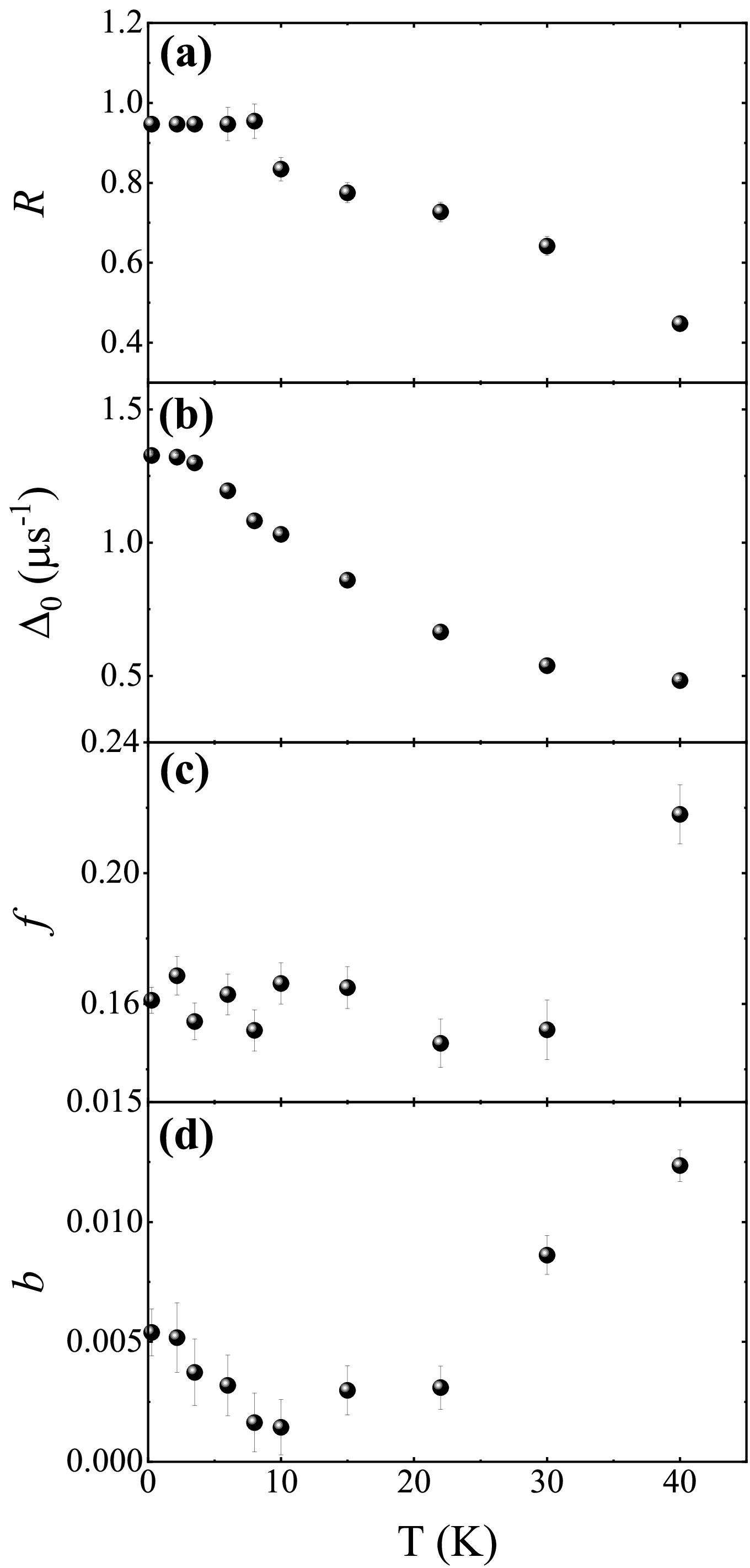}\\
  \caption{Temperature dependence of the parameters extracted from Eq. \ref{c}}\label{musr2}
\end{figure}

More insights into the spin dynamics of the title compound are obtained from local-probe $\mu$SR measurements. Figure \ref{musr}(a) summarizes the ZF-$\mu$SR spectra collected at various temperatures. At high temperatures (80 K), the $\mu$SR asymmetry shows a typical Kubo-Toyabe (KT) behavior with a dip around $\sim$6 $\mu$s and a recovery of the asymmetry at a longer time. This is typically observed in systems with randomly-oriented static internal fields with a Gaussian distribution due to the nuclear moments \cite{Guo2013}. The spectra can be well described by
\begin{equation}\label{a}
  A(t) = A_s\cdot\mathrm{KT}^{G}(t)\mathrm{exp}(-\lambda t) + b,
\end{equation}
where KT$^{G}$(\textit{t}) is the Kubo-Toyabe function with a Gaussian distribution
\begin{equation}\label{GKT}
  \mathrm{KT}^{G}(t) = \frac{1}{3} + \frac{2}{3}(1 - \Delta^2 t^2)\mathrm{exp}(-\frac{\Delta^2 t^2}{2}),
\end{equation}
and the exponential term represents additional contributions from the electronic spins. $\Delta/\gamma_\mu$ is the root-mean-square (rms) of the local-field distribution, and $\gamma_\mu$ = 2$\pi\times$13.55 MHz/kG is the gyromagnetic ratio of muons.
The best fit yields a small background, \textit{b}, of 0.011(1) compared to a large $A_s$ of 0.257(1), indicating that most of the muons are stopped at the sample position. The extracted  $\Delta/\gamma_\mu$ = 3.6 G and $\lambda$ = 0.048(5) $\mu$s$^{-1}$ suggest that the relaxation is mainly caused by the nuclear moments \cite{Guo2013}.
As the temperature decreases, the initial relaxation rate increases, while the dip becomes shallower than that expected from the KT$^{G}$(t) function. Below $\sim$6 K, the spectra are barely changed, and the dip is completely absent. Note, that the temperature scale (6 K) here is consistent with the broad peak observed in the specific heat. Moreover, the flat tail is larger than the background, and the initial relaxation is Gaussian-shaped instead of Lorentzian-shaped, usually expected for the dynamic electronic spins; see inset in Fig. \ref{musr}(a). These observations suggest that the local fields are static within the time window of $\mu$SR, which is further corroborated by the LF measurements shown in Fig. \ref{musr}(b,c). At the base temperature (0.27 K), the flat tail is gradually recovered with increasing LF and the asymmetry is fully recovered at 100 mT. A similar behavior is also observed at 40 K.

The shape of the ZF spectra is reminiscent of the Gaussian-broadened-Gaussian (GbG) function \cite{Noakes1997} in which the $\Delta$ in Eq. \ref{GKT} has a Gaussian distribution with a mean value of $\Delta_0$ and a rms of \textit{W}. The GbG(t) function is expressed as

\begin{equation}\label{b}
\begin{split}
  \mathrm{GbG}(t) = & f + (1-f)(\frac{1}{1 + R^2 \Delta_0^2t^2})^{3/2}(1-\frac{\Delta_0^2 t^2}{1 + R^2 \Delta_0^2 t^2}) \\
                    & \times\mathrm{exp}[-\frac{\Delta_0^2 t^2}{2(1 + R^2 \Delta_0^2 t^2)}],
\end{split}
\end{equation}
where $f$ = 1/3 for a perfect powder sample, and $R = W/\Delta_0$. Note that the tails in our spectra deviate from 1/3 of the total asymmetry, most likely because of the appearance of a preferred orientation since a pellet was used. In addition, we found a slight change of the parameter $\alpha$ at lower temperatures, probably due to a small change of the sample position at different temperatures. Since this small change of $\alpha$ only slightly shifts the spectra, we can describe them with:
\begin{equation}\label{c}
  A(t) = A_s\cdot\mathrm{GbG}(t) + b,
\end{equation}
where the amplitude $A_s$ was fixed to the value extracted from the 80 K spectrum, and $R$ is also fixed below 6 K.

The temperature dependence of the extracted parameters is shown in Fig. \ref{musr2}. Both \textit{R} and $\Delta_0$ increase monotonically as the temperature decreases, indicating that the system becomes more disordered at lower temperatures. The mean field strength at the muon site is estimated to be $\sqrt{8/\pi}$$\Delta_0/\gamma_\mu \sim$ 25 G at 0.27 K, which is much larger than that obtained from the high temperature spectrum, suggesting that its origin is closely related to the electron spins. The portion of the flat tail is already below the ``1/3" value at 40 K, and it does not vary much below 30 K, indicating that the electronic spins begin to set in at about 30 K.

The observation of static, disordered magnetism from the $\mu$SR measurement is quite surprising, since the dc and ac susceptibility data do not show any sign of freezing or anomaly either at around 30 K or 6 K. One possibility to reconcile this discrepancy may be derived from the sensitivity to different dynamic ranges of the different techniques. The $\mu$SR technique covers a time window of about 10$^{-12}$ $\rightarrow$ 10$^{-6}$ s \cite{Yaouanc2011}, while our ac susceptibility is restricted to the order of 10$^{-4}$ $\rightarrow$ 10$^{-1}$ s. Therefore, the spins fluctuating between the kHz to MHz range will behave as static from the viewpoint of $\mu$SR, but as dynamic for the ac susceptibility. These persistent slow fluctuations extend over about 2 orders of temperature range, from as high as 30 K down to 50 mK, demonstrating the strong correlations among the spins, while the quantum fluctuations still prevent the system from ordering or freezing down to mK range.

A more plausible origin could be the implanted muon induced modification of the local environment, which lowers the CEF symmetry and splits the ground state doublet into two singlets, that could facilitate the enhancement of Pr nuclear moments via hyperfine interactions, as observed in some Pr-pyrochlores \cite{MacLaughlin2009,Foronda2015}. In this case, the observed static magnetism stems from the nuclear moments, and the electron moments keep fluctuating from the view point of $\mu$SR.

\section{Conclusion}
In summary, we have probed the spin dynamics of PrZnAl$_{11}$O$_{19}$ down to 50 mK. AC susceptibility and $\mu$SR measurements indicate spin fluctuations down to 50 mK in spite of a large Curie-Weiss temperature. Low-energy magnetic excitations with a large density of states have been identified from ac susceptibility, heat capacity, and INS measurements. The gapless character of spin excitations in PrZnAl$_{11}$O$_{19}$ is verified by the power-law behavior of the heat capacity. All these suggest that PrZnAl$_{11}$O$_{19}$ is a good QSL candidate with a well-defined 2D triangular lattice. However, some details, such as the nontrivial field dependence of the excitations (as already revealed by the specific-heat measurement), the exact CEF ground state, and magnetic excitations at lower temperatures, need more theoretical and experimental elaborations based on single-crystal studies.

\begin{acknowledgments}

We thank J. Xu for helpful discussions. This work is supported by the NSF of China with Grant No. 12004270, 11874158 and 92065203, the Guangdong Basic and Applied Basic Research Foundation (2019A1515110517), and the Research Grants Council of Hong Kong with General Research Fund Grant No. 17306520. A portion of this work was supported by the Laboratory Directed Research and Development (LDRD) program of Oak Ridge National Laboratory, managed by UT-Battelle, LLC for the U.S. Department of Energy. We gratefully acknowledge the Science and Technology Facilities Council (STFC) for Xpress access to neutron beamtime on MERLIN at ISIS. Part of this work is based on experiments performed at the Swiss Muon Source S$\mu$S, Paul Scherrer Institute, Villigen, Switzerland.

\end{acknowledgments}

\bibliography{ref}

\begin{thebibliography}{50}%
\makeatletter
\providecommand \@ifxundefined [1]{%
 \@ifx{#1\undefined}
}%
\providecommand \@ifnum [1]{%
 \ifnum #1\expandafter \@firstoftwo
 \else \expandafter \@secondoftwo
 \fi
}%
\providecommand \@ifx [1]{%
 \ifx #1\expandafter \@firstoftwo
 \else \expandafter \@secondoftwo
 \fi
}%
\providecommand \natexlab [1]{#1}%
\providecommand \enquote  [1]{``#1''}%
\providecommand \bibnamefont  [1]{#1}%
\providecommand \bibfnamefont [1]{#1}%
\providecommand \citenamefont [1]{#1}%
\providecommand \href@noop [0]{\@secondoftwo}%
\providecommand \href [0]{\begingroup \@sanitize@url \@href}%
\providecommand \@href[1]{\@@startlink{#1}\@@href}%
\providecommand \@@href[1]{\endgroup#1\@@endlink}%
\providecommand \@sanitize@url [0]{\catcode `\\12\catcode `\$12\catcode
  `\&12\catcode `\#12\catcode `\^12\catcode `\_12\catcode `\%12\relax}%
\providecommand \@@startlink[1]{}%
\providecommand \@@endlink[0]{}%
\providecommand \url  [0]{\begingroup\@sanitize@url \@url }%
\providecommand \@url [1]{\endgroup\@href {#1}{\urlprefix }}%
\providecommand \urlprefix  [0]{URL }%
\providecommand \Eprint [0]{\href }%
\providecommand \doibase [0]{http://dx.doi.org/}%
\providecommand \selectlanguage [0]{\@gobble}%
\providecommand \bibinfo  [0]{\@secondoftwo}%
\providecommand \bibfield  [0]{\@secondoftwo}%
\providecommand \translation [1]{[#1]}%
\providecommand \BibitemOpen [0]{}%
\providecommand \bibitemStop [0]{}%
\providecommand \bibitemNoStop [0]{.\EOS\space}%
\providecommand \EOS [0]{\spacefactor3000\relax}%
\providecommand \BibitemShut  [1]{\csname bibitem#1\endcsname}%
\let\auto@bib@innerbib\@empty
\bibitem [{\citenamefont {Nayak}\ \emph {et~al.}(2008)\citenamefont {Nayak},
  \citenamefont {Simon}, \citenamefont {Stern}, \citenamefont {Freedman},\ and\
  \citenamefont {Das~Sarma}}]{Nayak2008}%
  \BibitemOpen
  \bibfield  {author} {\bibinfo {author} {\bibfnamefont {C.}~\bibnamefont
  {Nayak}}, \bibinfo {author} {\bibfnamefont {S.~H.}\ \bibnamefont {Simon}},
  \bibinfo {author} {\bibfnamefont {A.}~\bibnamefont {Stern}}, \bibinfo
  {author} {\bibfnamefont {M.}~\bibnamefont {Freedman}}, \ and\ \bibinfo
  {author} {\bibfnamefont {S.}~\bibnamefont {Das~Sarma}},\ }\href {\doibase
  10.1103/RevModPhys.80.1083} {\bibfield  {journal} {\bibinfo  {journal} {Rev.
  Mod. Phys.}\ }\textbf {\bibinfo {volume} {80}},\ \bibinfo {pages} {1083}
  (\bibinfo {year} {2008})}\BibitemShut {NoStop}%
\bibitem [{\citenamefont {Anderson}(1987)}]{Anderson1987}%
  \BibitemOpen
  \bibfield  {author} {\bibinfo {author} {\bibfnamefont {P.~W.}\ \bibnamefont
  {Anderson}},\ }\href {\doibase 10.1126/science.235.4793.1196} {\bibfield
  {journal} {\bibinfo  {journal} {Science}\ }\textbf {\bibinfo {volume}
  {235}},\ \bibinfo {pages} {1196} (\bibinfo {year} {1987})}\BibitemShut
  {NoStop}%
\bibitem [{\citenamefont {Anderson}(1973)}]{Anderson1973}%
  \BibitemOpen
  \bibfield  {author} {\bibinfo {author} {\bibfnamefont {P.~W.}\ \bibnamefont
  {Anderson}},\ }\href {\doibase https://doi.org/10.1016/0025-5408(73)90167-0}
  {\bibfield  {journal} {\bibinfo  {journal} {Mat. Res. Bull.}\ }\textbf
  {\bibinfo {volume} {8}},\ \bibinfo {pages} {153} (\bibinfo {year}
  {1973})}\BibitemShut {NoStop}%
\bibitem [{\citenamefont {Balents}(2010)}]{Balents2010}%
  \BibitemOpen
  \bibfield  {author} {\bibinfo {author} {\bibfnamefont {L.}~\bibnamefont
  {Balents}},\ }\href {\doibase 10.1038/nature08917} {\bibfield  {journal}
  {\bibinfo  {journal} {Nature}\ }\textbf {\bibinfo {volume} {464}},\ \bibinfo
  {pages} {199} (\bibinfo {year} {2010})}\BibitemShut {NoStop}%
\bibitem [{\citenamefont {Savary}\ and\ \citenamefont
  {Balents}(2016)}]{Savary2016}%
  \BibitemOpen
  \bibfield  {author} {\bibinfo {author} {\bibfnamefont {L.}~\bibnamefont
  {Savary}}\ and\ \bibinfo {author} {\bibfnamefont {L.}~\bibnamefont
  {Balents}},\ }\href {\doibase 10.1088/0034-4885/80/1/016502} {\bibfield
  {journal} {\bibinfo  {journal} {Rep. Prog. Phys.}\ }\textbf {\bibinfo
  {volume} {80}},\ \bibinfo {pages} {016502} (\bibinfo {year}
  {2016})}\BibitemShut {NoStop}%
\bibitem [{\citenamefont {Zhou}\ \emph {et~al.}(2017)\citenamefont {Zhou},
  \citenamefont {Kanoda},\ and\ \citenamefont {Ng}}]{Zhou2017}%
  \BibitemOpen
  \bibfield  {author} {\bibinfo {author} {\bibfnamefont {Y.}~\bibnamefont
  {Zhou}}, \bibinfo {author} {\bibfnamefont {K.}~\bibnamefont {Kanoda}}, \ and\
  \bibinfo {author} {\bibfnamefont {T.-K.}\ \bibnamefont {Ng}},\ }\href
  {\doibase 10.1103/RevModPhys.89.025003} {\bibfield  {journal} {\bibinfo
  {journal} {Rev. Mod. Phys.}\ }\textbf {\bibinfo {volume} {89}},\ \bibinfo
  {pages} {025003} (\bibinfo {year} {2017})}\BibitemShut {NoStop}%
\bibitem [{\citenamefont {Broholm}\ \emph {et~al.}(2020)\citenamefont
  {Broholm}, \citenamefont {Cava}, \citenamefont {Kivelson}, \citenamefont
  {Nocera}, \citenamefont {Norman},\ and\ \citenamefont
  {Senthil}}]{Broholm2020}%
  \BibitemOpen
  \bibfield  {author} {\bibinfo {author} {\bibfnamefont {C.}~\bibnamefont
  {Broholm}}, \bibinfo {author} {\bibfnamefont {R.~J.}\ \bibnamefont {Cava}},
  \bibinfo {author} {\bibfnamefont {S.~A.}\ \bibnamefont {Kivelson}}, \bibinfo
  {author} {\bibfnamefont {D.~G.}\ \bibnamefont {Nocera}}, \bibinfo {author}
  {\bibfnamefont {M.~R.}\ \bibnamefont {Norman}}, \ and\ \bibinfo {author}
  {\bibfnamefont {T.}~\bibnamefont {Senthil}},\ }\href {\doibase
  10.1126/science.aay0668} {\bibfield  {journal} {\bibinfo  {journal}
  {Science}\ }\textbf {\bibinfo {volume} {367}},\ \bibinfo {pages} {eaay0668}
  (\bibinfo {year} {2020})}\BibitemShut {NoStop}%
\bibitem [{\citenamefont {Wen}(2002)}]{Wen2002}%
  \BibitemOpen
  \bibfield  {author} {\bibinfo {author} {\bibfnamefont {X.-G.}\ \bibnamefont
  {Wen}},\ }\href {\doibase 10.1103/PhysRevB.65.165113} {\bibfield  {journal}
  {\bibinfo  {journal} {Phys. Rev. B}\ }\textbf {\bibinfo {volume} {65}},\
  \bibinfo {pages} {165113} (\bibinfo {year} {2002})}\BibitemShut {NoStop}%
\bibitem [{\citenamefont {Shimizu}\ \emph {et~al.}(2003)\citenamefont
  {Shimizu}, \citenamefont {Miyagawa}, \citenamefont {Kanoda}, \citenamefont
  {Maesato},\ and\ \citenamefont {Saito}}]{Shimizu2003}%
  \BibitemOpen
  \bibfield  {author} {\bibinfo {author} {\bibfnamefont {Y.}~\bibnamefont
  {Shimizu}}, \bibinfo {author} {\bibfnamefont {K.}~\bibnamefont {Miyagawa}},
  \bibinfo {author} {\bibfnamefont {K.}~\bibnamefont {Kanoda}}, \bibinfo
  {author} {\bibfnamefont {M.}~\bibnamefont {Maesato}}, \ and\ \bibinfo
  {author} {\bibfnamefont {G.}~\bibnamefont {Saito}},\ }\href {\doibase
  10.1103/PhysRevLett.91.107001} {\bibfield  {journal} {\bibinfo  {journal}
  {Phys. Rev. Lett.}\ }\textbf {\bibinfo {volume} {91}},\ \bibinfo {pages}
  {107001} (\bibinfo {year} {2003})}\BibitemShut {NoStop}%
\bibitem [{\citenamefont {Yamashita}\ \emph {et~al.}(2008)\citenamefont
  {Yamashita}, \citenamefont {Nakazawa}, \citenamefont {Oguni}, \citenamefont
  {Oshima}, \citenamefont {Nojiri}, \citenamefont {Shimizu}, \citenamefont
  {Miyagawa},\ and\ \citenamefont {Kanoda}}]{Yamashita2008}%
  \BibitemOpen
  \bibfield  {author} {\bibinfo {author} {\bibfnamefont {S.}~\bibnamefont
  {Yamashita}}, \bibinfo {author} {\bibfnamefont {Y.}~\bibnamefont {Nakazawa}},
  \bibinfo {author} {\bibfnamefont {M.}~\bibnamefont {Oguni}}, \bibinfo
  {author} {\bibfnamefont {Y.}~\bibnamefont {Oshima}}, \bibinfo {author}
  {\bibfnamefont {H.}~\bibnamefont {Nojiri}}, \bibinfo {author} {\bibfnamefont
  {Y.}~\bibnamefont {Shimizu}}, \bibinfo {author} {\bibfnamefont
  {K.}~\bibnamefont {Miyagawa}}, \ and\ \bibinfo {author} {\bibfnamefont
  {K.}~\bibnamefont {Kanoda}},\ }\href {\doibase 10.1038/nphys942} {\bibfield
  {journal} {\bibinfo  {journal} {Nat. Phys.}\ }\textbf {\bibinfo {volume}
  {4}},\ \bibinfo {pages} {459} (\bibinfo {year} {2008})}\BibitemShut {NoStop}%
\bibitem [{\citenamefont {Helton}\ \emph {et~al.}(2007)\citenamefont {Helton},
  \citenamefont {Matan}, \citenamefont {Shores}, \citenamefont {Nytko},
  \citenamefont {Bartlett}, \citenamefont {Yoshida}, \citenamefont {Takano},
  \citenamefont {Suslov}, \citenamefont {Qiu}, \citenamefont {Chung},
  \citenamefont {Nocera},\ and\ \citenamefont {Lee}}]{Helton2007}%
  \BibitemOpen
  \bibfield  {author} {\bibinfo {author} {\bibfnamefont {J.~S.}\ \bibnamefont
  {Helton}}, \bibinfo {author} {\bibfnamefont {K.}~\bibnamefont {Matan}},
  \bibinfo {author} {\bibfnamefont {M.~P.}\ \bibnamefont {Shores}}, \bibinfo
  {author} {\bibfnamefont {E.~A.}\ \bibnamefont {Nytko}}, \bibinfo {author}
  {\bibfnamefont {B.~M.}\ \bibnamefont {Bartlett}}, \bibinfo {author}
  {\bibfnamefont {Y.}~\bibnamefont {Yoshida}}, \bibinfo {author} {\bibfnamefont
  {Y.}~\bibnamefont {Takano}}, \bibinfo {author} {\bibfnamefont
  {A.}~\bibnamefont {Suslov}}, \bibinfo {author} {\bibfnamefont
  {Y.}~\bibnamefont {Qiu}}, \bibinfo {author} {\bibfnamefont {J.~H.}\
  \bibnamefont {Chung}}, \bibinfo {author} {\bibfnamefont {D.~G.}\ \bibnamefont
  {Nocera}}, \ and\ \bibinfo {author} {\bibfnamefont {Y.~S.}\ \bibnamefont
  {Lee}},\ }\href {\doibase 10.1103/PhysRevLett.98.107204} {\bibfield
  {journal} {\bibinfo  {journal} {Phys. Rev. Lett.}\ }\textbf {\bibinfo
  {volume} {98}},\ \bibinfo {pages} {107204} (\bibinfo {year}
  {2007})}\BibitemShut {NoStop}%
\bibitem [{\citenamefont {Gardner}\ \emph {et~al.}(1999)\citenamefont
  {Gardner}, \citenamefont {Dunsiger}, \citenamefont {Gaulin}, \citenamefont
  {Gingras}, \citenamefont {Greedan}, \citenamefont {Kiefl}, \citenamefont
  {Lumsden}, \citenamefont {MacFarlane}, \citenamefont {Raju}, \citenamefont
  {Sonier}, \citenamefont {Swainson},\ and\ \citenamefont {Tun}}]{Gardner1999}%
  \BibitemOpen
  \bibfield  {author} {\bibinfo {author} {\bibfnamefont {J.~S.}\ \bibnamefont
  {Gardner}}, \bibinfo {author} {\bibfnamefont {S.~R.}\ \bibnamefont
  {Dunsiger}}, \bibinfo {author} {\bibfnamefont {B.~D.}\ \bibnamefont
  {Gaulin}}, \bibinfo {author} {\bibfnamefont {M.~J.~P.}\ \bibnamefont
  {Gingras}}, \bibinfo {author} {\bibfnamefont {J.~E.}\ \bibnamefont
  {Greedan}}, \bibinfo {author} {\bibfnamefont {R.~F.}\ \bibnamefont {Kiefl}},
  \bibinfo {author} {\bibfnamefont {M.~D.}\ \bibnamefont {Lumsden}}, \bibinfo
  {author} {\bibfnamefont {W.~A.}\ \bibnamefont {MacFarlane}}, \bibinfo
  {author} {\bibfnamefont {N.~P.}\ \bibnamefont {Raju}}, \bibinfo {author}
  {\bibfnamefont {J.~E.}\ \bibnamefont {Sonier}}, \bibinfo {author}
  {\bibfnamefont {I.}~\bibnamefont {Swainson}}, \ and\ \bibinfo {author}
  {\bibfnamefont {Z.}~\bibnamefont {Tun}},\ }\href {\doibase
  10.1103/PhysRevLett.82.1012} {\bibfield  {journal} {\bibinfo  {journal}
  {Phys. Rev. Lett.}\ }\textbf {\bibinfo {volume} {82}},\ \bibinfo {pages}
  {1012} (\bibinfo {year} {1999})}\BibitemShut {NoStop}%
\bibitem [{\citenamefont {Ross}\ \emph {et~al.}(2011)\citenamefont {Ross},
  \citenamefont {Savary}, \citenamefont {Gaulin},\ and\ \citenamefont
  {Balents}}]{Ross2011}%
  \BibitemOpen
  \bibfield  {author} {\bibinfo {author} {\bibfnamefont {K.~A.}\ \bibnamefont
  {Ross}}, \bibinfo {author} {\bibfnamefont {L.}~\bibnamefont {Savary}},
  \bibinfo {author} {\bibfnamefont {B.~D.}\ \bibnamefont {Gaulin}}, \ and\
  \bibinfo {author} {\bibfnamefont {L.}~\bibnamefont {Balents}},\ }\href
  {\doibase 10.1103/PhysRevX.1.021002} {\bibfield  {journal} {\bibinfo
  {journal} {Phys. Rev. X}\ }\textbf {\bibinfo {volume} {1}},\ \bibinfo {pages}
  {021002} (\bibinfo {year} {2011})}\BibitemShut {NoStop}%
\bibitem [{\citenamefont {Kimura}\ \emph {et~al.}(2013)\citenamefont {Kimura},
  \citenamefont {Nakatsuji}, \citenamefont {Wen}, \citenamefont {Broholm},
  \citenamefont {Stone}, \citenamefont {Nishibori},\ and\ \citenamefont
  {Sawa}}]{Kimura2013}%
  \BibitemOpen
  \bibfield  {author} {\bibinfo {author} {\bibfnamefont {K.}~\bibnamefont
  {Kimura}}, \bibinfo {author} {\bibfnamefont {S.}~\bibnamefont {Nakatsuji}},
  \bibinfo {author} {\bibfnamefont {J.~J.}\ \bibnamefont {Wen}}, \bibinfo
  {author} {\bibfnamefont {C.}~\bibnamefont {Broholm}}, \bibinfo {author}
  {\bibfnamefont {M.~B.}\ \bibnamefont {Stone}}, \bibinfo {author}
  {\bibfnamefont {E.}~\bibnamefont {Nishibori}}, \ and\ \bibinfo {author}
  {\bibfnamefont {H.}~\bibnamefont {Sawa}},\ }\href {\doibase
  10.1038/ncomms2914} {\bibfield  {journal} {\bibinfo  {journal} {Nat.
  Commun.}\ }\textbf {\bibinfo {volume} {4}},\ \bibinfo {pages} {1934}
  (\bibinfo {year} {2013})}\BibitemShut {NoStop}%
\bibitem [{\citenamefont {Jackeli}\ and\ \citenamefont
  {Khaliullin}(2009)}]{Jackeli2009}%
  \BibitemOpen
  \bibfield  {author} {\bibinfo {author} {\bibfnamefont {G.}~\bibnamefont
  {Jackeli}}\ and\ \bibinfo {author} {\bibfnamefont {G.}~\bibnamefont
  {Khaliullin}},\ }\href {\doibase 10.1103/PhysRevLett.102.017205} {\bibfield
  {journal} {\bibinfo  {journal} {Phys. Rev. Lett.}\ }\textbf {\bibinfo
  {volume} {102}},\ \bibinfo {pages} {017205} (\bibinfo {year}
  {2009})}\BibitemShut {NoStop}%
\bibitem [{\citenamefont {Plumb}\ \emph {et~al.}(2014)\citenamefont {Plumb},
  \citenamefont {Clancy}, \citenamefont {Sandilands}, \citenamefont {Shankar},
  \citenamefont {Hu}, \citenamefont {Burch}, \citenamefont {Kee},\ and\
  \citenamefont {Kim}}]{Plumb2014}%
  \BibitemOpen
  \bibfield  {author} {\bibinfo {author} {\bibfnamefont {K.~W.}\ \bibnamefont
  {Plumb}}, \bibinfo {author} {\bibfnamefont {J.~P.}\ \bibnamefont {Clancy}},
  \bibinfo {author} {\bibfnamefont {L.~J.}\ \bibnamefont {Sandilands}},
  \bibinfo {author} {\bibfnamefont {V.~V.}\ \bibnamefont {Shankar}}, \bibinfo
  {author} {\bibfnamefont {Y.~F.}\ \bibnamefont {Hu}}, \bibinfo {author}
  {\bibfnamefont {K.~S.}\ \bibnamefont {Burch}}, \bibinfo {author}
  {\bibfnamefont {H.-Y.}\ \bibnamefont {Kee}}, \ and\ \bibinfo {author}
  {\bibfnamefont {Y.-J.}\ \bibnamefont {Kim}},\ }\href {\doibase
  10.1103/PhysRevB.90.041112} {\bibfield  {journal} {\bibinfo  {journal} {Phys.
  Rev. B}\ }\textbf {\bibinfo {volume} {90}},\ \bibinfo {pages} {041112}
  (\bibinfo {year} {2014})}\BibitemShut {NoStop}%
\bibitem [{\citenamefont {Banerjee}\ \emph {et~al.}(2016)\citenamefont
  {Banerjee}, \citenamefont {Bridges}, \citenamefont {Yan}, \citenamefont
  {Aczel}, \citenamefont {Li}, \citenamefont {Stone}, \citenamefont {Granroth},
  \citenamefont {Lumsden}, \citenamefont {Yiu}, \citenamefont {Knolle},
  \citenamefont {Bhattacharjee}, \citenamefont {Kovrizhin}, \citenamefont
  {Moessner}, \citenamefont {Tennant}, \citenamefont {Mandrus},\ and\
  \citenamefont {Nagler}}]{Banerjee2016}%
  \BibitemOpen
  \bibfield  {author} {\bibinfo {author} {\bibfnamefont {A.}~\bibnamefont
  {Banerjee}}, \bibinfo {author} {\bibfnamefont {C.~A.}\ \bibnamefont
  {Bridges}}, \bibinfo {author} {\bibfnamefont {J.~Q.}\ \bibnamefont {Yan}},
  \bibinfo {author} {\bibfnamefont {A.~A.}\ \bibnamefont {Aczel}}, \bibinfo
  {author} {\bibfnamefont {L.}~\bibnamefont {Li}}, \bibinfo {author}
  {\bibfnamefont {M.~B.}\ \bibnamefont {Stone}}, \bibinfo {author}
  {\bibfnamefont {G.~E.}\ \bibnamefont {Granroth}}, \bibinfo {author}
  {\bibfnamefont {M.~D.}\ \bibnamefont {Lumsden}}, \bibinfo {author}
  {\bibfnamefont {Y.}~\bibnamefont {Yiu}}, \bibinfo {author} {\bibfnamefont
  {J.}~\bibnamefont {Knolle}}, \bibinfo {author} {\bibfnamefont
  {S.}~\bibnamefont {Bhattacharjee}}, \bibinfo {author} {\bibfnamefont {D.~L.}\
  \bibnamefont {Kovrizhin}}, \bibinfo {author} {\bibfnamefont {R.}~\bibnamefont
  {Moessner}}, \bibinfo {author} {\bibfnamefont {D.~A.}\ \bibnamefont
  {Tennant}}, \bibinfo {author} {\bibfnamefont {D.~G.}\ \bibnamefont
  {Mandrus}}, \ and\ \bibinfo {author} {\bibfnamefont {S.~E.}\ \bibnamefont
  {Nagler}},\ }\href {\doibase 10.1038/nmat4604} {\bibfield  {journal}
  {\bibinfo  {journal} {Nat. Mater.}\ }\textbf {\bibinfo {volume} {15}},\
  \bibinfo {pages} {733} (\bibinfo {year} {2016})}\BibitemShut {NoStop}%
\bibitem [{\citenamefont {Li}\ \emph {et~al.}(2015)\citenamefont {Li},
  \citenamefont {Liao}, \citenamefont {Zhang}, \citenamefont {Li},
  \citenamefont {Jin}, \citenamefont {Ling}, \citenamefont {Zhang},
  \citenamefont {Zou}, \citenamefont {Pi}, \citenamefont {Yang}, \citenamefont
  {Wang}, \citenamefont {Wu},\ and\ \citenamefont {Zhang}}]{Li2015}%
  \BibitemOpen
  \bibfield  {author} {\bibinfo {author} {\bibfnamefont {Y.}~\bibnamefont
  {Li}}, \bibinfo {author} {\bibfnamefont {H.}~\bibnamefont {Liao}}, \bibinfo
  {author} {\bibfnamefont {Z.}~\bibnamefont {Zhang}}, \bibinfo {author}
  {\bibfnamefont {S.}~\bibnamefont {Li}}, \bibinfo {author} {\bibfnamefont
  {F.}~\bibnamefont {Jin}}, \bibinfo {author} {\bibfnamefont {L.}~\bibnamefont
  {Ling}}, \bibinfo {author} {\bibfnamefont {L.}~\bibnamefont {Zhang}},
  \bibinfo {author} {\bibfnamefont {Y.}~\bibnamefont {Zou}}, \bibinfo {author}
  {\bibfnamefont {L.}~\bibnamefont {Pi}}, \bibinfo {author} {\bibfnamefont
  {Z.}~\bibnamefont {Yang}}, \bibinfo {author} {\bibfnamefont {J.}~\bibnamefont
  {Wang}}, \bibinfo {author} {\bibfnamefont {Z.}~\bibnamefont {Wu}}, \ and\
  \bibinfo {author} {\bibfnamefont {Q.}~\bibnamefont {Zhang}},\ }\href
  {\doibase 10.1038/srep16419} {\bibfield  {journal} {\bibinfo  {journal} {Sci.
  Rep.}\ }\textbf {\bibinfo {volume} {5}},\ \bibinfo {pages} {16419} (\bibinfo
  {year} {2015})}\BibitemShut {NoStop}%
\bibitem [{\citenamefont {Li}\ \emph {et~al.}(2016)\citenamefont {Li},
  \citenamefont {Adroja}, \citenamefont {Biswas}, \citenamefont {Baker},
  \citenamefont {Zhang}, \citenamefont {Liu}, \citenamefont {Tsirlin},
  \citenamefont {Gegenwart},\ and\ \citenamefont {Zhang}}]{Li2016}%
  \BibitemOpen
  \bibfield  {author} {\bibinfo {author} {\bibfnamefont {Y.}~\bibnamefont
  {Li}}, \bibinfo {author} {\bibfnamefont {D.}~\bibnamefont {Adroja}}, \bibinfo
  {author} {\bibfnamefont {P.~K.}\ \bibnamefont {Biswas}}, \bibinfo {author}
  {\bibfnamefont {P.~J.}\ \bibnamefont {Baker}}, \bibinfo {author}
  {\bibfnamefont {Q.}~\bibnamefont {Zhang}}, \bibinfo {author} {\bibfnamefont
  {J.}~\bibnamefont {Liu}}, \bibinfo {author} {\bibfnamefont {A.~A.}\
  \bibnamefont {Tsirlin}}, \bibinfo {author} {\bibfnamefont {P.}~\bibnamefont
  {Gegenwart}}, \ and\ \bibinfo {author} {\bibfnamefont {Q.}~\bibnamefont
  {Zhang}},\ }\href {\doibase 10.1103/PhysRevLett.117.097201} {\bibfield
  {journal} {\bibinfo  {journal} {Phys. Rev. Lett.}\ }\textbf {\bibinfo
  {volume} {117}},\ \bibinfo {pages} {097201} (\bibinfo {year}
  {2016})}\BibitemShut {NoStop}%
\bibitem [{\citenamefont {Shen}\ \emph {et~al.}(2016)\citenamefont {Shen},
  \citenamefont {Li}, \citenamefont {Wo}, \citenamefont {Li}, \citenamefont
  {Shen}, \citenamefont {Pan}, \citenamefont {Wang}, \citenamefont {Walker},
  \citenamefont {Steffens}, \citenamefont {Boehm}, \citenamefont {Hao},
  \citenamefont {Quintero-Castro}, \citenamefont {Harriger}, \citenamefont
  {Frontzek}, \citenamefont {Hao}, \citenamefont {Meng}, \citenamefont {Zhang},
  \citenamefont {Chen},\ and\ \citenamefont {Zhao}}]{Shen2016}%
  \BibitemOpen
  \bibfield  {author} {\bibinfo {author} {\bibfnamefont {Y.}~\bibnamefont
  {Shen}}, \bibinfo {author} {\bibfnamefont {Y.-D.}\ \bibnamefont {Li}},
  \bibinfo {author} {\bibfnamefont {H.}~\bibnamefont {Wo}}, \bibinfo {author}
  {\bibfnamefont {Y.}~\bibnamefont {Li}}, \bibinfo {author} {\bibfnamefont
  {S.}~\bibnamefont {Shen}}, \bibinfo {author} {\bibfnamefont {B.}~\bibnamefont
  {Pan}}, \bibinfo {author} {\bibfnamefont {Q.}~\bibnamefont {Wang}}, \bibinfo
  {author} {\bibfnamefont {H.~C.}\ \bibnamefont {Walker}}, \bibinfo {author}
  {\bibfnamefont {P.}~\bibnamefont {Steffens}}, \bibinfo {author}
  {\bibfnamefont {M.}~\bibnamefont {Boehm}}, \bibinfo {author} {\bibfnamefont
  {Y.}~\bibnamefont {Hao}}, \bibinfo {author} {\bibfnamefont {D.~L.}\
  \bibnamefont {Quintero-Castro}}, \bibinfo {author} {\bibfnamefont {L.~W.}\
  \bibnamefont {Harriger}}, \bibinfo {author} {\bibfnamefont {M.~D.}\
  \bibnamefont {Frontzek}}, \bibinfo {author} {\bibfnamefont {L.}~\bibnamefont
  {Hao}}, \bibinfo {author} {\bibfnamefont {S.}~\bibnamefont {Meng}}, \bibinfo
  {author} {\bibfnamefont {Q.}~\bibnamefont {Zhang}}, \bibinfo {author}
  {\bibfnamefont {G.}~\bibnamefont {Chen}}, \ and\ \bibinfo {author}
  {\bibfnamefont {J.}~\bibnamefont {Zhao}},\ }\href {\doibase
  10.1038/nature20614} {\bibfield  {journal} {\bibinfo  {journal} {Nature}\
  }\textbf {\bibinfo {volume} {540}},\ \bibinfo {pages} {559} (\bibinfo {year}
  {2016})}\BibitemShut {NoStop}%
\bibitem [{\citenamefont {Shen}\ \emph {et~al.}(2018)\citenamefont {Shen},
  \citenamefont {Li}, \citenamefont {Walker}, \citenamefont {Steffens},
  \citenamefont {Boehm}, \citenamefont {Zhang}, \citenamefont {Shen},
  \citenamefont {Wo}, \citenamefont {Chen},\ and\ \citenamefont
  {Zhao}}]{Shen2018}%
  \BibitemOpen
  \bibfield  {author} {\bibinfo {author} {\bibfnamefont {Y.}~\bibnamefont
  {Shen}}, \bibinfo {author} {\bibfnamefont {Y.-D.}\ \bibnamefont {Li}},
  \bibinfo {author} {\bibfnamefont {H.~C.}\ \bibnamefont {Walker}}, \bibinfo
  {author} {\bibfnamefont {P.}~\bibnamefont {Steffens}}, \bibinfo {author}
  {\bibfnamefont {M.}~\bibnamefont {Boehm}}, \bibinfo {author} {\bibfnamefont
  {X.}~\bibnamefont {Zhang}}, \bibinfo {author} {\bibfnamefont
  {S.}~\bibnamefont {Shen}}, \bibinfo {author} {\bibfnamefont {H.}~\bibnamefont
  {Wo}}, \bibinfo {author} {\bibfnamefont {G.}~\bibnamefont {Chen}}, \ and\
  \bibinfo {author} {\bibfnamefont {J.}~\bibnamefont {Zhao}},\ }\href {\doibase
  10.1038/s41467-018-06588-1} {\bibfield  {journal} {\bibinfo  {journal} {Nat.
  Commun.}\ }\textbf {\bibinfo {volume} {9}},\ \bibinfo {pages} {4138}
  (\bibinfo {year} {2018})}\BibitemShut {NoStop}%
\bibitem [{\citenamefont {Ding}\ \emph {et~al.}(2019)\citenamefont {Ding},
  \citenamefont {Manuel}, \citenamefont {Bachus}, \citenamefont {Gru{\ss}ler},
  \citenamefont {Gegenwart}, \citenamefont {Singleton}, \citenamefont
  {Johnson}, \citenamefont {Walker}, \citenamefont {Adroja}, \citenamefont
  {Hillier},\ and\ \citenamefont {Tsirlin}}]{Ding2019}%
  \BibitemOpen
  \bibfield  {author} {\bibinfo {author} {\bibfnamefont {L.}~\bibnamefont
  {Ding}}, \bibinfo {author} {\bibfnamefont {P.}~\bibnamefont {Manuel}},
  \bibinfo {author} {\bibfnamefont {S.}~\bibnamefont {Bachus}}, \bibinfo
  {author} {\bibfnamefont {F.}~\bibnamefont {Gru{\ss}ler}}, \bibinfo {author}
  {\bibfnamefont {P.}~\bibnamefont {Gegenwart}}, \bibinfo {author}
  {\bibfnamefont {J.}~\bibnamefont {Singleton}}, \bibinfo {author}
  {\bibfnamefont {R.~D.}\ \bibnamefont {Johnson}}, \bibinfo {author}
  {\bibfnamefont {H.~C.}\ \bibnamefont {Walker}}, \bibinfo {author}
  {\bibfnamefont {D.~T.}\ \bibnamefont {Adroja}}, \bibinfo {author}
  {\bibfnamefont {A.~D.}\ \bibnamefont {Hillier}}, \ and\ \bibinfo {author}
  {\bibfnamefont {A.~A.}\ \bibnamefont {Tsirlin}},\ }\href {\doibase
  10.1103/PhysRevB.100.144432} {\bibfield  {journal} {\bibinfo  {journal}
  {Phys. Rev. B}\ }\textbf {\bibinfo {volume} {100}},\ \bibinfo {pages}
  {144432} (\bibinfo {year} {2019})}\BibitemShut {NoStop}%
\bibitem [{\citenamefont {Dai}\ \emph {et~al.}(2021)\citenamefont {Dai},
  \citenamefont {Zhang}, \citenamefont {Xie}, \citenamefont {Duan},
  \citenamefont {Gao}, \citenamefont {Zhu}, \citenamefont {Feng}, \citenamefont
  {Tao}, \citenamefont {Huang}, \citenamefont {Cao}, \citenamefont
  {Podlesnyak}, \citenamefont {Granroth}, \citenamefont {Everett},
  \citenamefont {Neuefeind}, \citenamefont {Voneshen}, \citenamefont {Wang},
  \citenamefont {Tan}, \citenamefont {Morosan}, \citenamefont {Wang},
  \citenamefont {Lin}, \citenamefont {Shu}, \citenamefont {Chen}, \citenamefont
  {Guo}, \citenamefont {Lu},\ and\ \citenamefont {Dai}}]{Dai2021}%
  \BibitemOpen
  \bibfield  {author} {\bibinfo {author} {\bibfnamefont {P.-L.}\ \bibnamefont
  {Dai}}, \bibinfo {author} {\bibfnamefont {G.}~\bibnamefont {Zhang}}, \bibinfo
  {author} {\bibfnamefont {Y.}~\bibnamefont {Xie}}, \bibinfo {author}
  {\bibfnamefont {C.}~\bibnamefont {Duan}}, \bibinfo {author} {\bibfnamefont
  {Y.}~\bibnamefont {Gao}}, \bibinfo {author} {\bibfnamefont {Z.}~\bibnamefont
  {Zhu}}, \bibinfo {author} {\bibfnamefont {E.}~\bibnamefont {Feng}}, \bibinfo
  {author} {\bibfnamefont {Z.}~\bibnamefont {Tao}}, \bibinfo {author}
  {\bibfnamefont {C.-L.}\ \bibnamefont {Huang}}, \bibinfo {author}
  {\bibfnamefont {H.}~\bibnamefont {Cao}}, \bibinfo {author} {\bibfnamefont
  {A.}~\bibnamefont {Podlesnyak}}, \bibinfo {author} {\bibfnamefont {G.~E.}\
  \bibnamefont {Granroth}}, \bibinfo {author} {\bibfnamefont {M.~S.}\
  \bibnamefont {Everett}}, \bibinfo {author} {\bibfnamefont {J.~C.}\
  \bibnamefont {Neuefeind}}, \bibinfo {author} {\bibfnamefont {D.}~\bibnamefont
  {Voneshen}}, \bibinfo {author} {\bibfnamefont {S.}~\bibnamefont {Wang}},
  \bibinfo {author} {\bibfnamefont {G.}~\bibnamefont {Tan}}, \bibinfo {author}
  {\bibfnamefont {E.}~\bibnamefont {Morosan}}, \bibinfo {author} {\bibfnamefont
  {X.}~\bibnamefont {Wang}}, \bibinfo {author} {\bibfnamefont {H.-Q.}\
  \bibnamefont {Lin}}, \bibinfo {author} {\bibfnamefont {L.}~\bibnamefont
  {Shu}}, \bibinfo {author} {\bibfnamefont {G.}~\bibnamefont {Chen}}, \bibinfo
  {author} {\bibfnamefont {Y.}~\bibnamefont {Guo}}, \bibinfo {author}
  {\bibfnamefont {X.}~\bibnamefont {Lu}}, \ and\ \bibinfo {author}
  {\bibfnamefont {P.}~\bibnamefont {Dai}},\ }\href {\doibase
  10.1103/PhysRevX.11.021044} {\bibfield  {journal} {\bibinfo  {journal} {Phys.
  Rev. X}\ }\textbf {\bibinfo {volume} {11}},\ \bibinfo {pages} {021044}
  (\bibinfo {year} {2021})}\BibitemShut {NoStop}%
\bibitem [{\citenamefont {Freedman}\ \emph {et~al.}(2010)\citenamefont
  {Freedman}, \citenamefont {Han}, \citenamefont {Prodi}, \citenamefont
  {M¨¹ller}, \citenamefont {Huang}, \citenamefont {Chen}, \citenamefont {Webb},
  \citenamefont {Lee}, \citenamefont {McQueen},\ and\ \citenamefont
  {Nocera}}]{Freedman2010}%
  \BibitemOpen
  \bibfield  {author} {\bibinfo {author} {\bibfnamefont {D.~E.}\ \bibnamefont
  {Freedman}}, \bibinfo {author} {\bibfnamefont {T.~H.}\ \bibnamefont {Han}},
  \bibinfo {author} {\bibfnamefont {A.}~\bibnamefont {Prodi}}, \bibinfo
  {author} {\bibfnamefont {P.}~\bibnamefont {M¨¹ller}}, \bibinfo {author}
  {\bibfnamefont {Q.-Z.}\ \bibnamefont {Huang}}, \bibinfo {author}
  {\bibfnamefont {Y.-S.}\ \bibnamefont {Chen}}, \bibinfo {author}
  {\bibfnamefont {S.~M.}\ \bibnamefont {Webb}}, \bibinfo {author}
  {\bibfnamefont {Y.~S.}\ \bibnamefont {Lee}}, \bibinfo {author} {\bibfnamefont
  {T.~M.}\ \bibnamefont {McQueen}}, \ and\ \bibinfo {author} {\bibfnamefont
  {D.~G.}\ \bibnamefont {Nocera}},\ }\href {\doibase 10.1021/ja1070398}
  {\bibfield  {journal} {\bibinfo  {journal} {J. Am. Chem. Soc.}\ }\textbf
  {\bibinfo {volume} {132}},\ \bibinfo {pages} {16185} (\bibinfo {year}
  {2010})}\BibitemShut {NoStop}%
\bibitem [{\citenamefont {Singh}(2010)}]{Singh2010}%
  \BibitemOpen
  \bibfield  {author} {\bibinfo {author} {\bibfnamefont {R.~R.~P.}\
  \bibnamefont {Singh}},\ }\href {\doibase 10.1103/PhysRevLett.104.177203}
  {\bibfield  {journal} {\bibinfo  {journal} {Phys. Rev. Lett.}\ }\textbf
  {\bibinfo {volume} {104}},\ \bibinfo {pages} {177203} (\bibinfo {year}
  {2010})}\BibitemShut {NoStop}%
\bibitem [{\citenamefont {Ma}\ \emph {et~al.}(2018)\citenamefont {Ma},
  \citenamefont {Wang}, \citenamefont {Dong}, \citenamefont {Zhang},
  \citenamefont {Li}, \citenamefont {Zheng}, \citenamefont {Yu}, \citenamefont
  {Wang}, \citenamefont {Che}, \citenamefont {Ran}, \citenamefont {Bao},
  \citenamefont {Cai}, \citenamefont {\v{C}erm\'{a}k}, \citenamefont
  {Schneidewind}, \citenamefont {Yano}, \citenamefont {Gardner}, \citenamefont
  {Lu}, \citenamefont {Yu}, \citenamefont {Liu}, \citenamefont {Li},
  \citenamefont {Li},\ and\ \citenamefont {Wen}}]{Ma2018}%
  \BibitemOpen
  \bibfield  {author} {\bibinfo {author} {\bibfnamefont {Z.}~\bibnamefont
  {Ma}}, \bibinfo {author} {\bibfnamefont {J.}~\bibnamefont {Wang}}, \bibinfo
  {author} {\bibfnamefont {Z.-Y.}\ \bibnamefont {Dong}}, \bibinfo {author}
  {\bibfnamefont {J.}~\bibnamefont {Zhang}}, \bibinfo {author} {\bibfnamefont
  {S.}~\bibnamefont {Li}}, \bibinfo {author} {\bibfnamefont {S.-H.}\
  \bibnamefont {Zheng}}, \bibinfo {author} {\bibfnamefont {Y.}~\bibnamefont
  {Yu}}, \bibinfo {author} {\bibfnamefont {W.}~\bibnamefont {Wang}}, \bibinfo
  {author} {\bibfnamefont {L.}~\bibnamefont {Che}}, \bibinfo {author}
  {\bibfnamefont {K.}~\bibnamefont {Ran}}, \bibinfo {author} {\bibfnamefont
  {S.}~\bibnamefont {Bao}}, \bibinfo {author} {\bibfnamefont {Z.}~\bibnamefont
  {Cai}}, \bibinfo {author} {\bibfnamefont {P.}~\bibnamefont {\v{C}erm\'{a}k}},
  \bibinfo {author} {\bibfnamefont {A.}~\bibnamefont {Schneidewind}}, \bibinfo
  {author} {\bibfnamefont {S.}~\bibnamefont {Yano}}, \bibinfo {author}
  {\bibfnamefont {J.~S.}\ \bibnamefont {Gardner}}, \bibinfo {author}
  {\bibfnamefont {X.}~\bibnamefont {Lu}}, \bibinfo {author} {\bibfnamefont
  {S.-L.}\ \bibnamefont {Yu}}, \bibinfo {author} {\bibfnamefont {J.-M.}\
  \bibnamefont {Liu}}, \bibinfo {author} {\bibfnamefont {S.}~\bibnamefont
  {Li}}, \bibinfo {author} {\bibfnamefont {J.-X.}\ \bibnamefont {Li}}, \ and\
  \bibinfo {author} {\bibfnamefont {J.}~\bibnamefont {Wen}},\ }\href {\doibase
  10.1103/PhysRevLett.120.087201} {\bibfield  {journal} {\bibinfo  {journal}
  {Phys. Rev. Lett.}\ }\textbf {\bibinfo {volume} {120}},\ \bibinfo {pages}
  {087201} (\bibinfo {year} {2018})}\BibitemShut {NoStop}%
\bibitem [{\citenamefont {Ross}\ \emph {et~al.}(2009)\citenamefont {Ross},
  \citenamefont {Ruff}, \citenamefont {Adams}, \citenamefont {Gardner},
  \citenamefont {Dabkowska}, \citenamefont {Qiu}, \citenamefont {Copley},\ and\
  \citenamefont {Gaulin}}]{Ross2009}%
  \BibitemOpen
  \bibfield  {author} {\bibinfo {author} {\bibfnamefont {K.~A.}\ \bibnamefont
  {Ross}}, \bibinfo {author} {\bibfnamefont {J.~P.~C.}\ \bibnamefont {Ruff}},
  \bibinfo {author} {\bibfnamefont {C.~P.}\ \bibnamefont {Adams}}, \bibinfo
  {author} {\bibfnamefont {J.~S.}\ \bibnamefont {Gardner}}, \bibinfo {author}
  {\bibfnamefont {H.~A.}\ \bibnamefont {Dabkowska}}, \bibinfo {author}
  {\bibfnamefont {Y.}~\bibnamefont {Qiu}}, \bibinfo {author} {\bibfnamefont
  {J.~R.~D.}\ \bibnamefont {Copley}}, \ and\ \bibinfo {author} {\bibfnamefont
  {B.~D.}\ \bibnamefont {Gaulin}},\ }\href {\doibase
  10.1103/PhysRevLett.103.227202} {\bibfield  {journal} {\bibinfo  {journal}
  {Phys. Rev. Lett.}\ }\textbf {\bibinfo {volume} {103}},\ \bibinfo {pages}
  {227202} (\bibinfo {year} {2009})}\BibitemShut {NoStop}%
\bibitem [{\citenamefont {Chang}\ \emph {et~al.}(2012)\citenamefont {Chang},
  \citenamefont {Onoda}, \citenamefont {Su}, \citenamefont {Kao}, \citenamefont
  {Tsuei}, \citenamefont {Yasui}, \citenamefont {Kakurai},\ and\ \citenamefont
  {Lees}}]{Chang2012}%
  \BibitemOpen
  \bibfield  {author} {\bibinfo {author} {\bibfnamefont {L.-J.}\ \bibnamefont
  {Chang}}, \bibinfo {author} {\bibfnamefont {S.}~\bibnamefont {Onoda}},
  \bibinfo {author} {\bibfnamefont {Y.}~\bibnamefont {Su}}, \bibinfo {author}
  {\bibfnamefont {Y.-J.}\ \bibnamefont {Kao}}, \bibinfo {author} {\bibfnamefont
  {K.-D.}\ \bibnamefont {Tsuei}}, \bibinfo {author} {\bibfnamefont
  {Y.}~\bibnamefont {Yasui}}, \bibinfo {author} {\bibfnamefont
  {K.}~\bibnamefont {Kakurai}}, \ and\ \bibinfo {author} {\bibfnamefont
  {M.~R.}\ \bibnamefont {Lees}},\ }\href {\doibase 10.1038/ncomms1989}
  {\bibfield  {journal} {\bibinfo  {journal} {Nat. Commun.}\ }\textbf {\bibinfo
  {volume} {3}},\ \bibinfo {pages} {992} (\bibinfo {year} {2012})}\BibitemShut
  {NoStop}%
\bibitem [{\citenamefont {Furukawa}\ \emph {et~al.}(2015)\citenamefont
  {Furukawa}, \citenamefont {Miyagawa}, \citenamefont {Itou}, \citenamefont
  {Ito}, \citenamefont {Taniguchi}, \citenamefont {Saito}, \citenamefont
  {Iguchi}, \citenamefont {Sasaki},\ and\ \citenamefont
  {Kanoda}}]{Furukawa2015}%
  \BibitemOpen
  \bibfield  {author} {\bibinfo {author} {\bibfnamefont {T.}~\bibnamefont
  {Furukawa}}, \bibinfo {author} {\bibfnamefont {K.}~\bibnamefont {Miyagawa}},
  \bibinfo {author} {\bibfnamefont {T.}~\bibnamefont {Itou}}, \bibinfo {author}
  {\bibfnamefont {M.}~\bibnamefont {Ito}}, \bibinfo {author} {\bibfnamefont
  {H.}~\bibnamefont {Taniguchi}}, \bibinfo {author} {\bibfnamefont
  {M.}~\bibnamefont {Saito}}, \bibinfo {author} {\bibfnamefont
  {S.}~\bibnamefont {Iguchi}}, \bibinfo {author} {\bibfnamefont
  {T.}~\bibnamefont {Sasaki}}, \ and\ \bibinfo {author} {\bibfnamefont
  {K.}~\bibnamefont {Kanoda}},\ }\href {\doibase
  10.1103/PhysRevLett.115.077001} {\bibfield  {journal} {\bibinfo  {journal}
  {Phys. Rev. Lett.}\ }\textbf {\bibinfo {volume} {115}},\ \bibinfo {pages}
  {077001} (\bibinfo {year} {2015})}\BibitemShut {NoStop}%
\bibitem [{\citenamefont {Ashtar}\ \emph {et~al.}(2019)\citenamefont {Ashtar},
  \citenamefont {Marwat}, \citenamefont {Gao}, \citenamefont {Zhang},
  \citenamefont {Pi}, \citenamefont {Yuan},\ and\ \citenamefont
  {Tian}}]{Ashtar2019}%
  \BibitemOpen
  \bibfield  {author} {\bibinfo {author} {\bibfnamefont {M.}~\bibnamefont
  {Ashtar}}, \bibinfo {author} {\bibfnamefont {M.~A.}\ \bibnamefont {Marwat}},
  \bibinfo {author} {\bibfnamefont {Y.~X.}\ \bibnamefont {Gao}}, \bibinfo
  {author} {\bibfnamefont {Z.~T.}\ \bibnamefont {Zhang}}, \bibinfo {author}
  {\bibfnamefont {L.}~\bibnamefont {Pi}}, \bibinfo {author} {\bibfnamefont
  {S.~L.}\ \bibnamefont {Yuan}}, \ and\ \bibinfo {author} {\bibfnamefont
  {Z.~M.}\ \bibnamefont {Tian}},\ }\href {\doibase 10.1039/C9TC02643F}
  {\bibfield  {journal} {\bibinfo  {journal} {J. Mater. Chem. C}\ }\textbf
  {\bibinfo {volume} {7}},\ \bibinfo {pages} {10073} (\bibinfo {year}
  {2019})}\BibitemShut {NoStop}%
\bibitem [{\citenamefont {Li}(2019)}]{Li2019}%
  \BibitemOpen
  \bibfield  {author} {\bibinfo {author} {\bibfnamefont {Y.}~\bibnamefont
  {Li}},\ }\href {\doibase https://doi.org/10.1002/qute.201900089} {\bibfield
  {journal} {\bibinfo  {journal} {Adv. Quantum Techno.}\ }\textbf {\bibinfo
  {volume} {2}},\ \bibinfo {pages} {1900089} (\bibinfo {year}
  {2019})}\BibitemShut {NoStop}%
\bibitem [{\citenamefont {et~al.}(2020)}]{data1}%
  \BibitemOpen
  \bibfield  {author} {\bibinfo {author} {\bibfnamefont {H.~G.}\ \bibnamefont
  {et~al.}},\ }\href@noop {} {\enquote {\bibinfo {title} {{CEF} ground state of
  quantum spin liquid candidates {REZ}n{A}l$_{11}${O}$_{19}$ ({RE} = {P}r{,}
  {N}d) {https://doi.org/10.5286/ISIS.E.RB1990296-1}},}\ } (\bibinfo {year}
  {2020})\BibitemShut {NoStop}%
\bibitem [{\citenamefont {et~al.}(2021)}]{data2}%
  \BibitemOpen
  \bibfield  {author} {\bibinfo {author} {\bibfnamefont {H.~G.}\ \bibnamefont
  {et~al.}},\ }\href@noop {} {\enquote {\bibinfo {title} {{P}honon measurement
  on the quantum spin liquid candidates {REZ}n{A}l$_{11}${O}$_{19}$ ({RE} =
  {P}r, {N}d) {https://doi.org/10.5286/ISIS.E.RB2190071-1}},}\ } (\bibinfo
  {year} {2021})\BibitemShut {NoStop}%
\bibitem [{\citenamefont {Bewley}\ \emph {et~al.}(2009)\citenamefont {Bewley},
  \citenamefont {Guidi},\ and\ \citenamefont {Bennington}}]{MERLIN}%
  \BibitemOpen
  \bibfield  {author} {\bibinfo {author} {\bibfnamefont {R.~I.}\ \bibnamefont
  {Bewley}}, \bibinfo {author} {\bibfnamefont {T.}~\bibnamefont {Guidi}}, \
  and\ \bibinfo {author} {\bibfnamefont {S.}~\bibnamefont {Bennington}},\
  }\href@noop {} {\bibfield  {journal} {\bibinfo  {journal} {Notiziario
  Neutroni e Luce di Sincrotrone}\ }\textbf {\bibinfo {volume} {14}},\ \bibinfo
  {pages} {22} (\bibinfo {year} {2009})}\BibitemShut {NoStop}%
\bibitem [{\citenamefont {Plumb}\ \emph {et~al.}(2019)\citenamefont {Plumb},
  \citenamefont {Changlani}, \citenamefont {Scheie}, \citenamefont {Zhang},
  \citenamefont {Krizan}, \citenamefont {Rodriguez-Rivera}, \citenamefont
  {Qiu}, \citenamefont {Winn}, \citenamefont {Cava},\ and\ \citenamefont
  {Broholm}}]{Plumb2019}%
  \BibitemOpen
  \bibfield  {author} {\bibinfo {author} {\bibfnamefont {K.~W.}\ \bibnamefont
  {Plumb}}, \bibinfo {author} {\bibfnamefont {H.~J.}\ \bibnamefont
  {Changlani}}, \bibinfo {author} {\bibfnamefont {A.}~\bibnamefont {Scheie}},
  \bibinfo {author} {\bibfnamefont {S.}~\bibnamefont {Zhang}}, \bibinfo
  {author} {\bibfnamefont {J.~W.}\ \bibnamefont {Krizan}}, \bibinfo {author}
  {\bibfnamefont {J.~A.}\ \bibnamefont {Rodriguez-Rivera}}, \bibinfo {author}
  {\bibfnamefont {Y.}~\bibnamefont {Qiu}}, \bibinfo {author} {\bibfnamefont
  {B.}~\bibnamefont {Winn}}, \bibinfo {author} {\bibfnamefont {R.~J.}\
  \bibnamefont {Cava}}, \ and\ \bibinfo {author} {\bibfnamefont {C.~L.}\
  \bibnamefont {Broholm}},\ }\href {\doibase 10.1038/s41567-018-0317-3}
  {\bibfield  {journal} {\bibinfo  {journal} {Nat. Phys.}\ }\textbf {\bibinfo
  {volume} {15}},\ \bibinfo {pages} {54} (\bibinfo {year} {2019})}\BibitemShut
  {NoStop}%
\bibitem [{\citenamefont {Bouvier}\ \emph {et~al.}(1991)\citenamefont
  {Bouvier}, \citenamefont {Lethuillier},\ and\ \citenamefont
  {Schmitt}}]{Bouvier1991}%
  \BibitemOpen
  \bibfield  {author} {\bibinfo {author} {\bibfnamefont {M.}~\bibnamefont
  {Bouvier}}, \bibinfo {author} {\bibfnamefont {P.}~\bibnamefont
  {Lethuillier}}, \ and\ \bibinfo {author} {\bibfnamefont {D.}~\bibnamefont
  {Schmitt}},\ }\href {\doibase 10.1103/PhysRevB.43.13137} {\bibfield
  {journal} {\bibinfo  {journal} {Phys. Rev. B}\ }\textbf {\bibinfo {volume}
  {43}},\ \bibinfo {pages} {13137} (\bibinfo {year} {1991})}\BibitemShut
  {NoStop}%
\bibitem [{\citenamefont {Nakatsuji}\ \emph {et~al.}(2005)\citenamefont
  {Nakatsuji}, \citenamefont {Nambu}, \citenamefont {Tonomura}, \citenamefont
  {Sakai}, \citenamefont {Jonas}, \citenamefont {Broholm}, \citenamefont
  {Tsunetsugu}, \citenamefont {Qiu},\ and\ \citenamefont
  {Maeno}}]{Nakatsuji2005}%
  \BibitemOpen
  \bibfield  {author} {\bibinfo {author} {\bibfnamefont {S.}~\bibnamefont
  {Nakatsuji}}, \bibinfo {author} {\bibfnamefont {Y.}~\bibnamefont {Nambu}},
  \bibinfo {author} {\bibfnamefont {H.}~\bibnamefont {Tonomura}}, \bibinfo
  {author} {\bibfnamefont {O.}~\bibnamefont {Sakai}}, \bibinfo {author}
  {\bibfnamefont {S.}~\bibnamefont {Jonas}}, \bibinfo {author} {\bibfnamefont
  {C.}~\bibnamefont {Broholm}}, \bibinfo {author} {\bibfnamefont
  {H.}~\bibnamefont {Tsunetsugu}}, \bibinfo {author} {\bibfnamefont
  {Y.}~\bibnamefont {Qiu}}, \ and\ \bibinfo {author} {\bibfnamefont
  {Y.}~\bibnamefont {Maeno}},\ }\href {\doibase doi:10.1126/science.1114727}
  {\bibfield  {journal} {\bibinfo  {journal} {Science}\ }\textbf {\bibinfo
  {volume} {309}},\ \bibinfo {pages} {1697} (\bibinfo {year}
  {2005})}\BibitemShut {NoStop}%
\bibitem [{\citenamefont {Podolsky}\ and\ \citenamefont
  {Kim}(2009)}]{Podolsky2009}%
  \BibitemOpen
  \bibfield  {author} {\bibinfo {author} {\bibfnamefont {D.}~\bibnamefont
  {Podolsky}}\ and\ \bibinfo {author} {\bibfnamefont {Y.~B.}\ \bibnamefont
  {Kim}},\ }\href {\doibase 10.1103/PhysRevB.79.140402} {\bibfield  {journal}
  {\bibinfo  {journal} {Phys. Rev. B}\ }\textbf {\bibinfo {volume} {79}},\
  \bibinfo {pages} {140402} (\bibinfo {year} {2009})}\BibitemShut {NoStop}%
\bibitem [{\citenamefont {Stoudenmire}\ \emph {et~al.}(2009)\citenamefont
  {Stoudenmire}, \citenamefont {Trebst},\ and\ \citenamefont
  {Balents}}]{Stoudenmire2009}%
  \BibitemOpen
  \bibfield  {author} {\bibinfo {author} {\bibfnamefont {E.~M.}\ \bibnamefont
  {Stoudenmire}}, \bibinfo {author} {\bibfnamefont {S.}~\bibnamefont {Trebst}},
  \ and\ \bibinfo {author} {\bibfnamefont {L.}~\bibnamefont {Balents}},\ }\href
  {\doibase 10.1103/PhysRevB.79.214436} {\bibfield  {journal} {\bibinfo
  {journal} {Phys. Rev. B}\ }\textbf {\bibinfo {volume} {79}},\ \bibinfo
  {pages} {214436} (\bibinfo {year} {2009})}\BibitemShut {NoStop}%
\bibitem [{\citenamefont {Li}\ \emph {et~al.}(2019)\citenamefont {Li},
  \citenamefont {Jin}, \citenamefont {Guo}, \citenamefont {Xu}, \citenamefont
  {Su}, \citenamefont {Feng}, \citenamefont {Liu}, \citenamefont {Zhou},
  \citenamefont {Ying}, \citenamefont {Li}, \citenamefont {Wang}, \citenamefont
  {Chen},\ and\ \citenamefont {Chen}}]{Li2019prb}%
  \BibitemOpen
  \bibfield  {author} {\bibinfo {author} {\bibfnamefont {K.}~\bibnamefont
  {Li}}, \bibinfo {author} {\bibfnamefont {S.}~\bibnamefont {Jin}}, \bibinfo
  {author} {\bibfnamefont {J.}~\bibnamefont {Guo}}, \bibinfo {author}
  {\bibfnamefont {Y.}~\bibnamefont {Xu}}, \bibinfo {author} {\bibfnamefont
  {Y.}~\bibnamefont {Su}}, \bibinfo {author} {\bibfnamefont {E.}~\bibnamefont
  {Feng}}, \bibinfo {author} {\bibfnamefont {Y.}~\bibnamefont {Liu}}, \bibinfo
  {author} {\bibfnamefont {S.}~\bibnamefont {Zhou}}, \bibinfo {author}
  {\bibfnamefont {T.}~\bibnamefont {Ying}}, \bibinfo {author} {\bibfnamefont
  {S.}~\bibnamefont {Li}}, \bibinfo {author} {\bibfnamefont {Z.}~\bibnamefont
  {Wang}}, \bibinfo {author} {\bibfnamefont {G.}~\bibnamefont {Chen}}, \ and\
  \bibinfo {author} {\bibfnamefont {X.}~\bibnamefont {Chen}},\ }\href {\doibase
  10.1103/PhysRevB.99.054421} {\bibfield  {journal} {\bibinfo  {journal} {Phys.
  Rev. B}\ }\textbf {\bibinfo {volume} {99}},\ \bibinfo {pages} {054421}
  (\bibinfo {year} {2019})}\BibitemShut {NoStop}%
\bibitem [{\citenamefont {Liu}\ \emph {et~al.}(2018)\citenamefont {Liu},
  \citenamefont {Li},\ and\ \citenamefont {Chen}}]{Liu2018}%
  \BibitemOpen
  \bibfield  {author} {\bibinfo {author} {\bibfnamefont {C.}~\bibnamefont
  {Liu}}, \bibinfo {author} {\bibfnamefont {Y.-D.}\ \bibnamefont {Li}}, \ and\
  \bibinfo {author} {\bibfnamefont {G.}~\bibnamefont {Chen}},\ }\href {\doibase
  10.1103/PhysRevB.98.045119} {\bibfield  {journal} {\bibinfo  {journal} {Phys.
  Rev. B}\ }\textbf {\bibinfo {volume} {98}},\ \bibinfo {pages} {045119}
  (\bibinfo {year} {2018})}\BibitemShut {NoStop}%
\bibitem [{\citenamefont {Ran}\ \emph {et~al.}(2007)\citenamefont {Ran},
  \citenamefont {Hermele}, \citenamefont {Lee},\ and\ \citenamefont
  {Wen}}]{Ran2007}%
  \BibitemOpen
  \bibfield  {author} {\bibinfo {author} {\bibfnamefont {Y.}~\bibnamefont
  {Ran}}, \bibinfo {author} {\bibfnamefont {M.}~\bibnamefont {Hermele}},
  \bibinfo {author} {\bibfnamefont {P.~A.}\ \bibnamefont {Lee}}, \ and\
  \bibinfo {author} {\bibfnamefont {X.-G.}\ \bibnamefont {Wen}},\ }\href
  {\doibase 10.1103/PhysRevLett.98.117205} {\bibfield  {journal} {\bibinfo
  {journal} {Phys. Rev. Lett.}\ }\textbf {\bibinfo {volume} {98}},\ \bibinfo
  {pages} {117205} (\bibinfo {year} {2007})}\BibitemShut {NoStop}%
\bibitem [{\citenamefont {Zeng}\ \emph {et~al.}(2022)\citenamefont {Zeng},
  \citenamefont {Ma}, \citenamefont {Wu}, \citenamefont {Li}, \citenamefont
  {Tao}, \citenamefont {Lu}, \citenamefont {Chen}, \citenamefont {Mi},
  \citenamefont {Song}, \citenamefont {Cao}, \citenamefont {Che}, \citenamefont
  {Li}, \citenamefont {Li}, \citenamefont {Luo}, \citenamefont {Meng},\ and\
  \citenamefont {Li}}]{Zeng2022}%
  \BibitemOpen
  \bibfield  {author} {\bibinfo {author} {\bibfnamefont {Z.}~\bibnamefont
  {Zeng}}, \bibinfo {author} {\bibfnamefont {X.}~\bibnamefont {Ma}}, \bibinfo
  {author} {\bibfnamefont {S.}~\bibnamefont {Wu}}, \bibinfo {author}
  {\bibfnamefont {H.-F.}\ \bibnamefont {Li}}, \bibinfo {author} {\bibfnamefont
  {Z.}~\bibnamefont {Tao}}, \bibinfo {author} {\bibfnamefont {X.}~\bibnamefont
  {Lu}}, \bibinfo {author} {\bibfnamefont {X.-h.}\ \bibnamefont {Chen}},
  \bibinfo {author} {\bibfnamefont {J.-X.}\ \bibnamefont {Mi}}, \bibinfo
  {author} {\bibfnamefont {S.-J.}\ \bibnamefont {Song}}, \bibinfo {author}
  {\bibfnamefont {G.-H.}\ \bibnamefont {Cao}}, \bibinfo {author} {\bibfnamefont
  {G.}~\bibnamefont {Che}}, \bibinfo {author} {\bibfnamefont {K.}~\bibnamefont
  {Li}}, \bibinfo {author} {\bibfnamefont {G.}~\bibnamefont {Li}}, \bibinfo
  {author} {\bibfnamefont {H.}~\bibnamefont {Luo}}, \bibinfo {author}
  {\bibfnamefont {Z.~Y.}\ \bibnamefont {Meng}}, \ and\ \bibinfo {author}
  {\bibfnamefont {S.}~\bibnamefont {Li}},\ }\href {\doibase
  10.1103/PhysRevB.105.L121109} {\bibfield  {journal} {\bibinfo  {journal}
  {Phys. Rev. B}\ }\textbf {\bibinfo {volume} {105}},\ \bibinfo {pages}
  {L121109} (\bibinfo {year} {2022})}\BibitemShut {NoStop}%
\bibitem [{\citenamefont {Motrunich}(2005)}]{Motrunich2005}%
  \BibitemOpen
  \bibfield  {author} {\bibinfo {author} {\bibfnamefont {O.~I.}\ \bibnamefont
  {Motrunich}},\ }\href {\doibase 10.1103/PhysRevB.72.045105} {\bibfield
  {journal} {\bibinfo  {journal} {Phys. Rev. B}\ }\textbf {\bibinfo {volume}
  {72}},\ \bibinfo {pages} {045105} (\bibinfo {year} {2005})}\BibitemShut
  {NoStop}%
\bibitem [{\citenamefont {McEwen}\ \emph {et~al.}(2006)\citenamefont {McEwen},
  \citenamefont {Jensen}, \citenamefont {Beirne}, \citenamefont {Allen},
  \citenamefont {Habicht}, \citenamefont {Adroja}, \citenamefont {Bewley},\
  and\ \citenamefont {Fort}}]{McEwen2006}%
  \BibitemOpen
  \bibfield  {author} {\bibinfo {author} {\bibfnamefont {K.~A.}\ \bibnamefont
  {McEwen}}, \bibinfo {author} {\bibfnamefont {J.}~\bibnamefont {Jensen}},
  \bibinfo {author} {\bibfnamefont {E.~D.}\ \bibnamefont {Beirne}}, \bibinfo
  {author} {\bibfnamefont {J.~P.}\ \bibnamefont {Allen}}, \bibinfo {author}
  {\bibfnamefont {K.}~\bibnamefont {Habicht}}, \bibinfo {author} {\bibfnamefont
  {D.~T.}\ \bibnamefont {Adroja}}, \bibinfo {author} {\bibfnamefont {R.~I.}\
  \bibnamefont {Bewley}}, \ and\ \bibinfo {author} {\bibfnamefont
  {D.}~\bibnamefont {Fort}},\ }\href {\doibase 10.1103/PhysRevB.73.014402}
  {\bibfield  {journal} {\bibinfo  {journal} {Phys. Rev. B}\ }\textbf {\bibinfo
  {volume} {73}},\ \bibinfo {pages} {014402} (\bibinfo {year}
  {2006})}\BibitemShut {NoStop}%
\bibitem [{\citenamefont {Guo}\ \emph {et~al.}(2013)\citenamefont {Guo},
  \citenamefont {Tanida}, \citenamefont {Kobayashi}, \citenamefont {Kawasaki},
  \citenamefont {Sera}, \citenamefont {Nishioka}, \citenamefont {Matsumura},
  \citenamefont {Watanabe},\ and\ \citenamefont {Xu}}]{Guo2013}%
  \BibitemOpen
  \bibfield  {author} {\bibinfo {author} {\bibfnamefont {H.}~\bibnamefont
  {Guo}}, \bibinfo {author} {\bibfnamefont {H.}~\bibnamefont {Tanida}},
  \bibinfo {author} {\bibfnamefont {R.}~\bibnamefont {Kobayashi}}, \bibinfo
  {author} {\bibfnamefont {I.}~\bibnamefont {Kawasaki}}, \bibinfo {author}
  {\bibfnamefont {M.}~\bibnamefont {Sera}}, \bibinfo {author} {\bibfnamefont
  {T.}~\bibnamefont {Nishioka}}, \bibinfo {author} {\bibfnamefont
  {M.}~\bibnamefont {Matsumura}}, \bibinfo {author} {\bibfnamefont
  {I.}~\bibnamefont {Watanabe}}, \ and\ \bibinfo {author} {\bibfnamefont
  {Z.-a.}\ \bibnamefont {Xu}},\ }\href {\doibase 10.1103/PhysRevB.88.115206}
  {\bibfield  {journal} {\bibinfo  {journal} {Phys. Rev. B}\ }\textbf {\bibinfo
  {volume} {88}},\ \bibinfo {pages} {115206} (\bibinfo {year}
  {2013})}\BibitemShut {NoStop}%
\bibitem [{\citenamefont {Noakes}\ and\ \citenamefont
  {Kalvius}(1997)}]{Noakes1997}%
  \BibitemOpen
  \bibfield  {author} {\bibinfo {author} {\bibfnamefont {D.~R.}\ \bibnamefont
  {Noakes}}\ and\ \bibinfo {author} {\bibfnamefont {G.~M.}\ \bibnamefont
  {Kalvius}},\ }\href {\doibase 10.1103/PhysRevB.56.2352} {\bibfield  {journal}
  {\bibinfo  {journal} {Phys. Rev. B}\ }\textbf {\bibinfo {volume} {56}},\
  \bibinfo {pages} {2352} (\bibinfo {year} {1997})}\BibitemShut {NoStop}%
\bibitem [{\citenamefont {Yaouanc}\ and\ \citenamefont {Dalmas~de
  R\'{e}otier}(2011)}]{Yaouanc2011}%
  \BibitemOpen
  \bibfield  {author} {\bibinfo {author} {\bibfnamefont {A.}~\bibnamefont
  {Yaouanc}}\ and\ \bibinfo {author} {\bibfnamefont {P.}~\bibnamefont
  {Dalmas~de R\'{e}otier}},\ }\href@noop {} {\emph {\bibinfo {title} {{Muon
  Spin Rotation, Relaxation, and Resonance}}}}\ (\bibinfo  {publisher} {Oxford
  University Press},\ \bibinfo {year} {2011})\BibitemShut {NoStop}%
\bibitem [{\citenamefont {MacLaughlin}\ \emph {et~al.}(2009)\citenamefont
  {MacLaughlin}, \citenamefont {Ohta}, \citenamefont {Machida}, \citenamefont
  {Nakatsuji}, \citenamefont {Luke}, \citenamefont {Ishida}, \citenamefont
  {Heffner}, \citenamefont {Shu},\ and\ \citenamefont
  {Bernal}}]{MacLaughlin2009}%
  \BibitemOpen
  \bibfield  {author} {\bibinfo {author} {\bibfnamefont {D.~E.}\ \bibnamefont
  {MacLaughlin}}, \bibinfo {author} {\bibfnamefont {Y.}~\bibnamefont {Ohta}},
  \bibinfo {author} {\bibfnamefont {Y.}~\bibnamefont {Machida}}, \bibinfo
  {author} {\bibfnamefont {S.}~\bibnamefont {Nakatsuji}}, \bibinfo {author}
  {\bibfnamefont {G.~M.}\ \bibnamefont {Luke}}, \bibinfo {author}
  {\bibfnamefont {K.}~\bibnamefont {Ishida}}, \bibinfo {author} {\bibfnamefont
  {R.~H.}\ \bibnamefont {Heffner}}, \bibinfo {author} {\bibfnamefont
  {L.}~\bibnamefont {Shu}}, \ and\ \bibinfo {author} {\bibfnamefont {O.~O.}\
  \bibnamefont {Bernal}},\ }\href
  {http://www.sciencedirect.com/science/article/pii/S0921452608006376}
  {\bibfield  {journal} {\bibinfo  {journal} {Physica B}\ }\textbf {\bibinfo
  {volume} {404}},\ \bibinfo {pages} {667} (\bibinfo {year}
  {2009})}\BibitemShut {NoStop}%
\bibitem [{\citenamefont {Foronda}\ \emph {et~al.}(2015)\citenamefont
  {Foronda}, \citenamefont {Lang}, \citenamefont {M\"oller}, \citenamefont
  {Lancaster}, \citenamefont {Boothroyd}, \citenamefont {Pratt}, \citenamefont
  {Giblin}, \citenamefont {Prabhakaran},\ and\ \citenamefont
  {Blundell}}]{Foronda2015}%
  \BibitemOpen
  \bibfield  {author} {\bibinfo {author} {\bibfnamefont {F.~R.}\ \bibnamefont
  {Foronda}}, \bibinfo {author} {\bibfnamefont {F.}~\bibnamefont {Lang}},
  \bibinfo {author} {\bibfnamefont {J.~S.}\ \bibnamefont {M\"oller}}, \bibinfo
  {author} {\bibfnamefont {T.}~\bibnamefont {Lancaster}}, \bibinfo {author}
  {\bibfnamefont {A.~T.}\ \bibnamefont {Boothroyd}}, \bibinfo {author}
  {\bibfnamefont {F.~L.}\ \bibnamefont {Pratt}}, \bibinfo {author}
  {\bibfnamefont {S.~R.}\ \bibnamefont {Giblin}}, \bibinfo {author}
  {\bibfnamefont {D.}~\bibnamefont {Prabhakaran}}, \ and\ \bibinfo {author}
  {\bibfnamefont {S.~J.}\ \bibnamefont {Blundell}},\ }\href {\doibase
  10.1103/PhysRevLett.114.017602} {\bibfield  {journal} {\bibinfo  {journal}
  {Phys. Rev. Lett.}\ }\textbf {\bibinfo {volume} {114}},\ \bibinfo {pages}
  {017602} (\bibinfo {year} {2015})}\BibitemShut {NoStop}%
\end{thebibliography}%

\end{document}